\documentstyle[prb,aps]{revtex}
\begin{document}

\title{Improved Theory of Single-Particle Properties of Fermi Systems 
 Using Sea-Bosons}

\author{Girish S. Setlur}
\address{Department of Physics and Materials Research Laboratory,\\
 University of Illinois at Urbana-Champaign , Urbana Il 61801}
\maketitle

\begin{abstract}
In this article we combine the ideas introduced by us earlier in various
proportions to arrive at a simple and yet powerful means of studying
single-particle properties of homogeneous Fermi systems in detail without
making assumptions regarding the validity or otherwise of Fermi-liquid theory.
Novelties include the exact forms of the momentum distribution and spectral
functions in the intermediate density regime and a dielectric function that
is sensitive to significant qualitative changes in single-particle properties.
This time, care is taken to include the particle-hole mode by interpreting the
sum over modes as a weighted sum with the dynamical structure factor appearing
as the weight.
\end{abstract}

\hspace{13in}

\section{Introduction}

 Both the Hartree-Fock\cite{Hartree}
 and the Random Phase approximations(RPA)\cite{Bohm}
 are widely used in the physics literature to study Fermi systems.
 The Bogoliubov theory\cite{Bogoliubov}
 is the analog of the random-phase approximation for Bose systems.
 This latter fact has been
 demonstrated in detail in our article\cite{Setlur}.
 The analysis presented here could be repeated for Bose systems as well.
 But let us focus on the more important Fermi system.
 Here we try and address the
 question of validity of the Hartree-Fock approximation and the RPA.
 This is important since we showed in our earlier work that the simplest
 and most natural application of bosonization was in the regime where the
 RPA was exact\cite{Setlur}.
 It is therefore quite pertinent to ask whether this
 approximation is adequate. It is well-known that there are many problems
 with the RPA. In particular, the pair-correlation function has unphysical
 behaviour at short distances(it becomes negative)
 \cite{Mahan}. Therefore it is necessary to find
 better approximations. Many attempts have been made in the literature in
 this regard. The Hubbard approximation\cite{Hubbard}, the Singwi-Sjolander
 approach\cite{Singwi} and many others discussed in the text by Mahan
 \cite{Mahan} are among the more prominent. Many of them have found to 
 resolve this problem of short distance behavior of the pair correlation
 function. However it seems that none of them are able to address the
 question of single-particle properties.
 This issue has acquired an urgency in
 the recent past given that a segment of the high $ T_{c} $ community is 
 convinced that the non-superconducting state of these materials is 
 non-Fermi liquid. Therefore being able to find a theory that describes the
 non-Fermi liquid regime accurately is desirable. In this attempt, we 
 shall adopt a two-track approach. On the one hand we reinterpret the RPA 
 so that it becomes more widely applicable. 
 We argue that while the mean-field
 approximation carried out on the density operator
 is the Hartree-Fock approximation,
 the RPA manifests itself as mean-field theory applied not
 to the density operator but to the number operator. The density operator
 measures how the electrons are distributed in real space whereas the number
 operator measures how the electrons are distributed in momentum space. 
 Just as the Hartree-Fock approximation
 is valid when fluctuations in the density
 of electrons at each point in real space is small, the RPA is valid when
 fluctuations in the momentum distribution of electrons
 is small compared with the average momentum distribution
 which measures the probability of an electron to possess 
 a given momentum. We also introduce another mathematical transformation
 within the RPA-scheme and find that this maps a purely repulsive
 interaction to a purely attractive one but one that makes the particle exchange
 a momentum different from the usual one involved in two-body collisions.
 This unusual transformation allows us to define and compute yet another
 dielectric function of this system. A physical interpretation of this curious
 change in sign is given. A suitable combination of the purely repulsive(both actually
 and apparently) and purely attractive(only apparently, the electrons still
 repel one another) is shown to be better than either one spearately.
 In the end, we point out how a motivated researcher may be able to compute
 the phase diagram of the homogeneous electron gas by appropriately combining 
 these three ingredients, although we ourselves restrict our attention to
 simple analytically solvable cases.
\newline
\newline
(i) Sea-bosons, indispensable for computing single-particle properties.
\newline
\newline
(ii) Generalised RPA and beyond (with fluctuations in the momentum
 distribution). Many new dielectric functions may be found here.
\newline
\newline
(iii) Repulsion-attraction duality : the enigmatic transformation
 that allows us to view the purely repulsive interaction between electrons
 as being attractive and repulsive at the same time when viewed in the
 sea-boson language.
\newline
\newline 
 In what follows, we try and combine the bosonization approach of our
 earlier articles with this notion of generalised RPA and
 repulsion-attraction duality and try and
 compute the single-particle properties. Just in order to whet the reader's
 appetite for the calculation of single-particle properties using sea-bosons,
 we pause here to explore another option, one that suggests itself
 quite naturally when one views RPA as a mean-field idea applied to the number
 operator rather than the density operator.

\section{Single-Particle Properties Without Using Bosonization}

 Let us try and write down some representative examples when RPA and 
 Hartree-Fock approximations are used.
 Let us take for example, the jellium model \cite{Mahan}. 
\begin{equation}
H = \sum_{ {\bf{k}} }\epsilon_{ {\bf{k}} }c^{\dagger}_{ {\bf{k}} }c_{ {\bf{k}} }
+ \sum_{ {\bf{q}} \neq 0 }
\frac{ v_{ {\bf{q}} } }{2V}(\rho_{ {\bf{q}} }\rho_{ -{\bf{q}} } - N)
\end{equation}
 If we apply the mean-field idea on the density, that is, replace
 $ \rho_{ {\bf{q}} } $ by $ \langle \rho_{ {\bf{q}} } \rangle $
 then we get a hamiltonian that does not involve any coulomb interaction
 at all(apart from an additive constant).
\[
H = \sum_{ {\bf{k}} }\epsilon_{ {\bf{k}} }c^{\dagger}_{ {\bf{k}} }c_{ {\bf{k}} }
+ \sum_{ {\bf{q}} \neq 0 }
\frac{v_{ {\bf{q}} } }{2V}
(\langle \rho_{ {\bf{q}} } \rangle\rho_{ -{\bf{q}} } - N)
\]
\begin{equation}
= \sum_{ {\bf{k}} }\epsilon_{ {\bf{k}} }c^{\dagger}_{ {\bf{k}} }c_{ {\bf{k}} }
-N \sum_{ {\bf{q}} \neq 0 }
\frac{v_{ {\bf{q}} } }{2V} 
\end{equation}
 Therefore this approximation is bad in the extreme. However the
 random-phase approximation is still valid for this system.
 It is the mean-field idea applied not
 to the density but to the number operator.
 For this we have to rewrite the full hamiltonian given below,
\begin{equation}
H = \sum_{ {\bf{k}} }\epsilon_{ {\bf{k}} }c^{\dagger}_{ {\bf{k}} }c_{ {\bf{k}} }
+ \sum_{ {\bf{q}} \neq 0 }
\frac{ v_{ {\bf{q}} } }{2V}\sum_{ {\bf{k}} , {\bf{k}}^{'} }
c^{\dagger}_{ {\bf{k}}+{\bf{q}}/2 }c^{\dagger}_{ {\bf{k}}^{'}-{\bf{q}}/2 }
c_{ {\bf{k}}^{'}+{\bf{q}}/2 }c_{ {\bf{k}}-{\bf{q}}/2 }
\end{equation}
This has to be replaced by,
\begin{equation}
H = 
H_{0}
- \sum_{ {\bf{q}} \neq 0 }
\frac{v_{ {\bf{q}} } }{2V}\sum_{ {\bf{k}} \neq  {\bf{k}}^{'} }
c^{\dagger}_{ {\bf{k}}+{\bf{q}}/2 }
c_{ {\bf{k}}^{'}+{\bf{q}}/2 }
 c^{\dagger}_{ {\bf{k}}^{'}-{\bf{q}}/2 }c_{ {\bf{k}}-{\bf{q}}/2 } 
\end{equation}
where
\begin{equation}
H_{0} = \sum_{ {\bf{k}} }
\epsilon_{ {\bf{k}} }c^{\dagger}_{ {\bf{k}} }c_{ {\bf{k}} }
- \sum_{ {\bf{q}} \neq 0 }
\frac{ v_{ {\bf{q}} } }{2V}\sum_{ {\bf{k}} }
n_{ {\bf{k}}+{\bf{q}}/2 } n_{ {\bf{k}}-{\bf{q}}/2 }
\end{equation}
 where $ n_{ {\bf{k}} } = c^{\dagger}_{ {\bf{k}} }c_{ {\bf{k}} } $.
 Let us write,
\begin{equation}
 n_{ {\bf{k}} } = \langle n_{ {\bf{k}} } \rangle
  + \delta  n_{ {\bf{k}} } 
\end{equation}
 We plug the above decomposition into $ H_{0} $ and find that 
 if we neglect terms quadratic in the fluctuations, we get what
 we are after, namely the generalised RPA called for simplicity as just RPA.
\begin{equation}
H_{RPA} = E^{'}_{0} + \sum_{ {\bf{k}} }{\tilde{ \epsilon }}_{ {\bf{k}} }
c^{\dagger}_{ {\bf{k}} }c_{ {\bf{k}} }
\end{equation}
\begin{equation}
{\tilde{ \epsilon }}_{ {\bf{k}} } = \epsilon_{ {\bf{k}} }
 - \sum_{ {\bf{q}} }\frac{ v_{ {\bf{q}} } }{V} \langle n_{ {\bf{k-q}} } \rangle
\end{equation}
 The average occupation is
\begin{equation}
\langle n_{ {\bf{k}} } \rangle = \frac{1}
{exp(\beta({\tilde{ \epsilon }}_{ {\bf{k}} } -\mu)) + 1}
\label{SELF1}
\end{equation}
The chemical potential $ \mu $ has to be fixed by making sure that,
\begin{equation}
 \sum_{ {\bf{k}} } \langle n_{ {\bf{k}} } \rangle  = \langle N \rangle
\end{equation}
 At zero temperature $ \mu = \epsilon_{F} $,
 this quantity is equal to the usual Fermi energy
 when $ v_{ {\bf{q}} } = 0 $.
\begin{equation}
  \langle n_{ {\bf{k}} } \rangle = \theta(\epsilon_{F}
-{\tilde{ \epsilon }}_{ {\bf{k}} }) =  \theta(\epsilon_{F}
- \epsilon_{ {\bf{k}} }+ \sum_{ {\bf{q}} \neq 0 }
\frac{ v_{ {\bf{q}} } }{V}\langle n_{ {\bf{k}}-{\bf{q}} } \rangle)
\label{SELF2}
\end{equation}
 and $ \theta $ is the Heaviside step function.
 We can now demonstrate that the
 generalised RPA dielectric function(the Lindhard dielectric function
 \cite{Lind} being a weak coupling limit of this)
 may be recovered using the following procedure.
 If one considers an extremely weak external perturbation applied to
 the system and
 follows the discussion in Mahan \cite{Mahan}
 one arrives at the following formula for the dielectric function,
\begin{equation}
\epsilon_{g-RPA}({\bf{q}},\omega) = 
 1 + \frac{ v_{ {\bf{q}} } }{V}\sum_{ {\bf{k}} }
\frac{ \langle n_{ {\bf{k+q/2}} } \rangle - \langle  n_{ {\bf{k-q/2}} } \rangle }
{ \omega - {\tilde{\epsilon}}_{ {\bf{k+q/2}} }
 + {\tilde{\epsilon}}_{ {\bf{k-q/2}} } }
\label{GENRPA}
\end{equation}
 The only point to bear in mind is that we have to use the full interacting
 momentum distribution rather than just its noninteracting value. When one
 includes fluctuations in the momentum distribution however, 
 the answer given later, is very different from the usual
 RPA or even the generalised RPA. 
 This feature of having the full momentum distribution
 in the numerator may be found in our earlier work \cite{Setlur}.
 Thus we can see
 that there is a whole new set of approximations that go beyond the RPA.
 While none of these revelations may come as a surprise to the reader,
 it should serve as a reminder that even our most cherished approximations
 may not be controlled in any sense of the term.
 It is more likely that
 they were the first to appear in the literature 
 and probably the easiest ones to use thus explaining their popularity. 
 The dielectric function written down above has the attractive feature of
 reducing to the familiar Lindhard dielectric function\cite{Lind}
 for extremely weak
 coupling and at the same time giving us something very different for
 stronger coupling. Furthermore, if we ask the Luttinger liquid community to
 write down a dielectric function of the Luttinger liquid, they are in
 all probability, going to write down the traditional RPA
 dielectric function\cite{Neto}. But we submit that
 this is a serious mistake. The more nonideal a system is, the more its
 dielectric function differs from the traditional RPA, even at high density.
 Parenthetically, we note that having granted the fact that the system is
 nonideal(namely, no Fermi surface or close to being devoid of one)
 it makes no difference whether the system is at high or low density,
 the point is, it is the {\it{generalised}} RPA that is still valid
 for this system.
 The only drawback of the above approach is that if we
 compute the one-particle Green functions we find that the imaginary
 part vanishes identically. This is unfortunate and we have to do better
 in order capture lifetime effects. The RPA hamiltonian neglects 
 fluctuations in the momentum distribution of the electrons. In order
 to recover a finite lifetime of single particle excitations,
 we find that it is important to  study the generalised $ H_{0} $
 rather than the more simple $ H_{RPA} $.
 The fluctuations in the momentum
 distribution may be related to the mean by the following observation.
 Define,
\begin{equation} 
 N({\bf{k}}, {\bf{k}}^{'})
 = \langle n_{ {\bf{k}} }n_{ {\bf{k}}^{'} } \rangle -
 \langle n_{ {\bf{k}} } \rangle  \langle n_{ {\bf{k}}^{'} } \rangle 
\end{equation}
 . The fluctuation in the number operator is
 $ N({\bf{k}}, {\bf{k}}) =  \langle n^{2}_{ {\bf{k}} } \rangle - \langle n_{ {\bf{k}} } \rangle^{2} $ .
 Since $ n^{2}_{ {\bf{k}} } = n_{ {\bf{k}} }  $ for fermions, we have
\begin{equation} 
 N({\bf{k}}, {\bf{k}})
 = \langle n_{ {\bf{k}} } \rangle(1-\langle n_{ {\bf{k}} } \rangle) 
\label{INTEGRALEQ}
\end{equation}  
 Therefore, we may conclude that any nonideal momentum distribution fluctuates
 (there are however, pathological exceptions, see footnote\cite{footnote}).
 In fact, a very nonideal momentum distribution such as one for which 
 $ \langle n_{ {\bf{k}} }  \rangle = 0.5 $ for most momenta has
 the largest fluctuation. When dealing with nonideal
 systems, we are obliged to consider fluctuations in the momentum
 distribution. In order to study lifetime effects therefore, we have to include
 the full $ H_{0} $. This may be written more transparently as
 (apart from additive constants) $ H _{0} = H_{RPA} + H_{fl} $,
\begin{equation}
H_{fl} = -\sum_{ {\bf{q}} \neq 0 }\frac{ v_{ {\bf{q}} } }{2V}
\sum_{ {\bf{k}} } \delta n_{ {\bf{k}} + {\bf{q}}/2 } 
\delta n_{ {\bf{k}} - {\bf{q}}/2 }
\end{equation}
 The full Fermi propagator may be evaluated
 by treating the fluctuation part as a perturbation and using the
 functional methods of Schwinger illustrated brilliantly by Kadanoff and
 Baym\cite{Baym}. 
 Before we plunge into the details it is important to keep in mind
 that the mean also changes when we consider fluctuations. That is,
\begin{equation}
\langle n_{ {\bf{k}} } \rangle = \langle n_{ {\bf{k}} } \rangle_{RPA}
 + \langle n_{ {\bf{k}} } \rangle_{fl}
\end{equation}
 here $ \langle n_{ {\bf{k}} } \rangle_{RPA} $ is given by
 Eq.(~\ref{SELF1}) or Eq.(~\ref{SELF2}) the rest is nonzero only
 when the fluctuation in the momentum distribution is large
 (which is, unfortunately, almost always the case when
 $ \langle n_{ {\bf{k}} } \rangle_{RPA} $ is nonideal)
 Therefore, now, $ \delta n_{ {\bf{k}} } $ refers to fluctuation around
 the full average rather than the RPA average.
 The final answers are given below, and it is hoped that the reader can
 rederive them using the references quoted in the bibliography
 (mainly \cite{Mahan}, \cite{Baym}).
 We shall adhere to the notation of Kadanoff and Baym \cite{Baym}.
 In their notation the final answers for the single-particle
 Green functions are as follows(we assume in the following that
 $ F({\bf{p}}) \neq 0 $, however one may investigate the limit
 $  F({\bf{p}}) \rightarrow 0 $, here $ z_{n} = (2n+1)\pi/\beta $),
\begin{equation}
G_{n}({\bf{k}}) = \frac{1}{iz_{n} - {\tilde{\epsilon}}_{ {\bf{k}} } + \mu
 - \Sigma_{n}({\bf{k}}) }
\end{equation}
and 
\begin{equation}
\Sigma_{n}({\bf{k}}) = G_{n}({\bf{k}})F({\bf{k}})
\label{SELF}
\end{equation} 
\begin{equation}
F({\bf{k}}) = \sum_{ {\bf{q}},{\bf{q}}^{'} \neq 0 }
\frac{ v_{ {\bf{q}} }v_{ {\bf{q}}^{'} } }{V^{2}}
N({\bf{k-q}},{\bf{k}}-{\bf{q}}^{'})
\end{equation}
 From Eq.(~\ref{SELF}) we may obtain the real and imaginary parts of the
 retarded self-energy, and from there the spectral function and the 
 collision rates\cite{Baym}.
\begin{equation}
\Gamma({\bf{p}},\omega) = \sqrt{-\kappa({\bf{p}},\omega)}
\end{equation}
 Similarly,
\begin{equation}
A({\bf{p}},\omega) = \frac{ \sqrt{-\kappa({\bf{p}},\omega)} }{F({\bf{p}})}
\end{equation}
 if $ \kappa({\bf{p}},\omega) = (\omega-{\tilde{\epsilon}}_{ {\bf{p}} }+\mu)^{2}-4F({\bf{p}}) < 0 $
 and both are zero otherwise(when $ F({\bf{p}}) \neq 0 $).
 It can be seen that the spectral function is peaked around 
 $ {\tilde{\epsilon}}_{ {\bf{p}} } - \mu $ with a width of the order
 of $ 2\sqrt{ F({\bf{p}}) } $, and the collision
 rate is vanishingly small for those values of $ {\bf{p}} $
 for which $ F({\bf{p}}) $ is close to zero.
 It may be shown that the momentum
 distribution including possible fluctuations is given by,
\begin{equation}
\langle n_{ {\bf{p}} } \rangle = (\frac{2}{\pi})
\int^{\pi/2}_{-\pi/2}d\theta\mbox{       }cos^{2}\theta\mbox{         }
\frac{1}{e^{\beta({\tilde{\epsilon}}_{ {\bf{p}} }-\mu)}
e^{2\beta \sqrt{ F({\bf{p}}) }sin \theta}
 + 1}
\label{FLUCMOM}
\end{equation}
 and $ \langle n_{ {\bf{p}} } \rangle = 0 $ for $ F({\bf{p}}) < 0 $.
 Again, it may be seen quite easily that Eq.(~\ref{FLUCMOM}) is identical to
 Eq.(~\ref{SELF1}) when fluctuations in the momentum distribution
 are ignored.
 The only sticking point now is the computation of the fluctuation in the
 momentum distribution. This may be done in a similar manner by employing
 the functional methods of Kadanoff and Baym\cite{Baym}. However, we shall
 defer the computation of this quantity until the next section. Incidentally,
 the formulas we suggested in our preprints for this quantity in retrospect,
 are probably rather poor. Instead we shall adopt a more rigorous
 approach in the next section.
 The dielectric function is also modified as a result of these fluctuations 
 and the final formula for the dielectric function including
 possible fluctuations in the momentum distribution is,
\begin{equation}
\epsilon_{eff}({\bf{q}},\omega) = \epsilon_{g-RPA}({\bf{q}},\omega)
 - (\frac{v_{ {\bf{q}} } }{V})^{2}\frac{ P_{2}({\bf{q}},\omega) }
{ \epsilon_{g-RPA}({\bf{q}},\omega) }
\label{ZEROS}
\end{equation}
Here,
\begin{equation}
P_{2}({\bf{q}},\omega) = \sum_{ {\bf{k}},{\bf{k}}^{'} }
\frac{ N({\bf{k}}+{\bf{q}}/2, {\bf{k}}^{'}+{\bf{q}}/2)
- N({\bf{k}}-{\bf{q}}/2, {\bf{k}}^{'}+{\bf{q}}/2)
- N({\bf{k}}+{\bf{q}}/2, {\bf{k}}^{'}-{\bf{q}}/2)
+ N({\bf{k}}-{\bf{q}}/2, {\bf{k}}^{'}-{\bf{q}}/2) }
{ (\omega - {\tilde{\epsilon}}_{ {\bf{k}}-{\bf{q}}/2 } 
+ {\tilde{\epsilon}}_{ {\bf{k}}+{\bf{q}}/2 })
(\omega - {\tilde{\epsilon}}_{ {\bf{k}}^{'}-{\bf{q}}/2 }
 + {\tilde{\epsilon}}_{ {\bf{k}}^{'}+{\bf{q}}/2 }) }
\end{equation}
 A full derivation of this is relegated to the appendix in order to 
 avoid interupting the flow of ideas. Assuming that $ \omega $ has a
 small imaginary part, we recover both the real and imaginary parts of
 the full dielectric function. 

\section{Repulsion-Attraction Duality}

In this section, we point out an interesting and even a seemingly 
paradoxical duality between attraction and repulsion. In a
homogeneous electron gas, electron repel each other. However, we may show
that by virtue of the electron being fermions,
an exchange operation on the interaction
part of the full hamiltonian leads to a change in sign of the interaction
and also a change in the momentum exchanged by the interacting electrons.
To see this, let us write down the interaction part of the full 
hamiltonian again.
\begin{equation}
H_{int} = 
 \sum_{ {\bf{q}} \neq 0 }
\frac{ v_{ {\bf{q}} } }{2V}\sum_{ {\bf{k}} \neq {\bf{k}}^{'} }
c^{\dagger}_{ {\bf{k}}+{\bf{q}}/2 }c^{\dagger}_{ {\bf{k}}^{'}-{\bf{q}}/2 }
c_{ {\bf{k}}^{'}+{\bf{q}}/2 }c_{ {\bf{k}}-{\bf{q}}/2 }
\end{equation}
 We may rearrange the various fermion operators in this interaction so that
 precisely the same quantity may be rewritten as,
\begin{equation}
H_{int} = 
 -\sum_{ {\bf{k}} \neq {\bf{k}}^{'} }
\frac{ v_{ {\bf{k}}-{\bf{k}}^{'} } }{2V}\sum_{ {\bf{q}} \neq 0 }
c^{\dagger}_{ {\bf{k}}+{\bf{q}}/2 }c^{\dagger}_{ {\bf{k}}^{'}-{\bf{q}}/2 }
c_{ {\bf{k}}^{'}+{\bf{q}}/2 }c_{ {\bf{k}}-{\bf{q}}/2 }
\end{equation}
 Two important changes have occured. First there is the change in sign.
 This is the reason for the term repulsion-attraction duality.
 The second is 
 the momentum that is exchanged is different. The momentum carried away by
 the virtual photon is no longer $ {\bf{q}} $ but is $ {\bf{k}}-{\bf{k}}^{'} $.
 The change in sign is peculiar to fermions, whereas the other feature is
 present for bosons too. Let us now use this new form and recompute the
 dielectric function. Actually, we could have done the same shuffle for
 the $ {\bf{k}} = {\bf{k}}^{'} $ part of the interaction. But we shall 
 relegate a more careful examination of these issues to future publications.
 The main purpose of the present article is to lay before the reader a
 scheme that is robust, rich and sufficiently general so that she may
 apply the ideas to other more practical problems. 
 In order to compute the dielectric function 
 suggested by this shuffled interaction let us proceed as follows.
 The full hamiltonian may be written as
($ n_{0}({\bf{k}}) = n_{ {\bf{k}} } = c^{\dagger}_{ {\bf{k}} }c_{ {\bf{k}} }$),
\[
H = \sum_{ {\bf{k}} }{\tilde{\epsilon}}_{ {\bf{k}} }n_{0}({\bf{k}})
 - \sum_{ {\bf{q}} \neq 0 }\frac{ v({\bf{q}}) }{2V}
\sum_{ {\bf{k}} }\delta n_{0}( {\bf{k}} + {\bf{q}}/2 )
\mbox{  }\delta n_{0}( {\bf{k}} - {\bf{q}}/2 )
\]
\begin{equation}
 -\sum_{ {\bf{k}} \neq {\bf{k}}^{'} }
\frac{ v_{ {\bf{k}}-{\bf{k}}^{'} } }{2V}\sum_{ {\bf{q}} \neq 0 }
n_{ {\bf{q}} }({\bf{k}})n_{ -{\bf{q}} }({\bf{k}}^{'})
\end{equation}
here $ n_{ {\bf{q}} }({\bf{k}}) = c^{\dagger}_{ {\bf{k}} + {\bf{q}}/2 }c_{ {\bf{k}} - {\bf{q}}/2 } $
Let us now apply an external field, 
\begin{equation}
H_{ext}(t) = \sum_{ {\bf{k}}, {\bf{q}} }
( U_{ext}({\bf{q}},t) + U^{*}_{ext}(-{\bf{q}},t) )
n_{ {\bf{q}} }({\bf{k}})
\end{equation}
 In order to proceed, we appeal to the random-phase approximation.
 In a suitably generalised sense, it may be defined to mean the following
 approximation :
\[
[n_{ {\bf{q}} }({\bf{k}}), n_{ {\bf{q}}^{'} }({\bf{k}}^{'})]
 = c^{\dagger}_{ {\bf{k}} + {\bf{q}}/2 }c_{ {\bf{k}}^{'} - {\bf{q}}^{'}/2 }
\delta_{ {\bf{k}} - {\bf{q}}/2 , {\bf{k}}^{'} + {\bf{q}}^{'}/2 }
 - c^{\dagger}_{ {\bf{k}}^{'} + {\bf{q}}^{'}/2 }
c_{ {\bf{k}} - {\bf{q}}/2 }
\delta_{ {\bf{k}}^{'} - {\bf{q}}^{'}/2 , {\bf{k}} + {\bf{q}}/2 }
\]
\begin{equation}
\approx \delta_{ {\bf{k}}, {\bf{k}}^{'} } \delta_{ {\bf{q}}, -{\bf{q}}^{'} }
(n_{0}({\bf{k}} + {\bf{q}}/2) - n_{0}({\bf{k}}- {\bf{q}}/2))
\end{equation}
\begin{equation}
[n_{ {\bf{q}} }({\bf{k}}), n_{0}({\bf{p}})] = 
 n_{ {\bf{q}} }({\bf{k}})(\delta_{ {\bf{p}}, {\bf{k}} - {\bf{q}}/2 }
 - \delta_{ {\bf{p}}, {\bf{k}} + {\bf{q}}/2 })
\end{equation}
 Here we have to retain $ n_{0}({\bf{k}}) $ as an operator in order to include
 effects due to fluctuations in the momentum distribution. Let us now
 write down the equation of motion of this operator.
\[
i\frac{ \partial  }{\partial t}
n_{ {\bf{q}} }({\bf{k}}) = ({\tilde{\epsilon}}_{ {\bf{k}} - {\bf{q}}/2 }
 - {\tilde{\epsilon}}_{ {\bf{k}} + {\bf{q}}/2 })
n_{ {\bf{q}} }({\bf{k}})
 - \sum_{ {\bf{q}}^{'} \neq 0 }\frac{ v({\bf{q}}^{'}) }{V}
(\delta n_{0}( {\bf{k}}-{\bf{q}}/2 - {\bf{q}}^{'} )
n_{ {\bf{q}} }({\bf{k}})
\]
\[
 - \delta n_{0}( {\bf{k}}+{\bf{q}}/2 - {\bf{q}}^{'} )n_{ {\bf{q}} }({\bf{k}}))
 -\sum_{ {\bf{k}} \neq {\bf{k}}^{'} }
\frac{ v_{ {\bf{k}}-{\bf{k}}^{'} } }{V}
(n_{0}({\bf{k}} + {\bf{q}}/2)n_{ {\bf{q}} }({\bf{k}}^{'})
 - n_{0}({\bf{k}} - {\bf{q}}/2)n_{ {\bf{q}} }({\bf{k}}^{'}))
\]
\begin{equation}
+ ( U_{ext}(-{\bf{q}},t) + U^{*}_{ext}({\bf{q}},t) )
 ( n_{0}({\bf{k}} + {\bf{q}}/2) 
 - n_{0}({\bf{k}} - {\bf{q}}/2) )
\end{equation}
Write $ n_{0}({\bf{k}}) = \langle n_{0}({\bf{k}}) \rangle + \delta\mbox{  } n_{0}({\bf{k}}) $
\[
i\frac{ \partial }{\partial t}
\langle n_{ {\bf{q}} }({\bf{k}}) \rangle
 = ({\tilde{\epsilon}}_{ {\bf{k}} - {\bf{q}}/2 }
 - {\tilde{\epsilon}}_{ {\bf{k}} + {\bf{q}}/2 })
\langle n_{ {\bf{q}} }({\bf{k}}) \rangle
\]
\[
 - \sum_{ {\bf{q}}^{'} \neq 0 }\frac{ v({\bf{q}}^{'}) }{V}
(\langle \delta n_{0}( {\bf{k}}-{\bf{q}}/2 - {\bf{q}}^{'} )
n_{ {\bf{q}} }({\bf{k}}) \rangle
 - \langle \delta n_{0}( {\bf{k}}+{\bf{q}}/2 - {\bf{q}}^{'} )
n_{ {\bf{q}} }({\bf{k}})\rangle )
\]
\[
 -\sum_{ {\bf{k}} \neq {\bf{k}}^{'} }
\frac{ v_{ {\bf{k}}-{\bf{k}}^{'} } }{V}
(\langle n_{0}({\bf{k}} + {\bf{q}}/2) \rangle
 - \langle n_{0}({\bf{k}} - {\bf{q}}/2) \rangle) 
 \langle n_{ {\bf{q}} }({\bf{k}}^{'}) \rangle 
\]
\[
 -\sum_{ {\bf{k}} \neq {\bf{k}}^{'} }
\frac{ v_{ {\bf{k}}-{\bf{k}}^{'} } }{V}
(\langle \delta n_{0}({\bf{k}} + {\bf{q}}/2)
 n_{ {\bf{q}} }({\bf{k}}^{'}) \rangle
 - \langle \delta n_{0}({\bf{k}} - {\bf{q}}/2)
 n_{ {\bf{q}} }({\bf{k}}^{'}) \rangle )
\]
\begin{equation}
+ ( U_{ext}(-{\bf{q}},t) + U^{*}_{ext}({\bf{q}},t) )
 ( \langle n_{0}({\bf{k}} + {\bf{q}}/2) \rangle 
 - \langle n_{0}({\bf{k}} - {\bf{q}}/2) \rangle )
\end{equation}

\[
i\frac{ \partial  }{\partial t}
\langle \delta n_{0}({\bf{p}})n_{ {\bf{q}} }({\bf{k}}) \rangle
 = ({\tilde{\epsilon}}_{ {\bf{k}} - {\bf{q}}/2 }
 - {\tilde{\epsilon}}_{ {\bf{k}} + {\bf{q}}/2 })
\langle \delta n_{0}({\bf{p}})n_{ {\bf{q}} }({\bf{k}}) \rangle
\]
\[
 - \sum_{ {\bf{q}}^{'} \neq 0 }\frac{ v({\bf{q}}^{'}) }{V}
(\langle \delta n_{0}({\bf{p}})
\delta n_{0}( {\bf{k}}-{\bf{q}}/2 - {\bf{q}}^{'} ) \rangle 
 - \langle \delta n_{0}({\bf{p}})
 \delta n_{0}( {\bf{k}}+{\bf{q}}/2 - {\bf{q}}^{'} ) \rangle)
\langle n_{ {\bf{q}} }({\bf{k}}) \rangle
\]
\[
 -\sum_{ {\bf{k}} \neq {\bf{k}}^{'} }
\frac{ v_{ {\bf{k}}-{\bf{k}}^{'} } }{V}
(\langle n_{0}({\bf{k}} + {\bf{q}}/2) \rangle - 
  \langle n_{0}({\bf{k}} - {\bf{q}}/2) \rangle)
 \langle \delta n_{0}({\bf{p}}) n_{ {\bf{q}} }({\bf{k}}^{'}) \rangle
\]
\[
 -\sum_{ {\bf{k}} \neq {\bf{k}}^{'} }
\frac{ v_{ {\bf{k}}-{\bf{k}}^{'} } }{V}
(\langle\delta n_{0}({\bf{p}})  \delta n_{0}({\bf{k}} + {\bf{q}}/2) \rangle
 - \langle \delta n_{0}({\bf{p}}) 
 \delta n_{0}({\bf{k}} - {\bf{q}}/2) \rangle)
 \langle n_{ {\bf{q}} }({\bf{k}}^{'}) \rangle
\]
\begin{equation}
+ ( U_{ext}(-{\bf{q}},t) + U^{*}_{ext}({\bf{q}},t) )
 ( \langle \delta n_{0}({\bf{p}})\delta 
 n_{0}({\bf{k}} + {\bf{q}}/2) \rangle 
 - \langle \delta n_{0}({\bf{p}})\delta
 n_{0}({\bf{k}} - {\bf{q}}/2) \rangle )
\end{equation}
Assuming that,
\begin{equation}
U_{ext}(-{\bf{q}},t) = U_{ext}(-{\bf{q}},0)e^{-i\omega \mbox{  }t}
\end{equation}
\[
(\omega - {\tilde{\epsilon}}_{ {\bf{k}} - {\bf{q}}/2 } 
 + {\tilde{\epsilon}}_{ {\bf{k}} + {\bf{q}}/2 })
\langle \mbox{  }
\delta \mbox{  }n_{0}({\bf{p}})\mbox{  }n_{ {\bf{q}} }({\bf{k}}) \rangle
 = -\sum_{ {\bf{q}}^{'} \neq 0 }\frac{ v_{ {\bf{q}}^{'} } }{V}
(N({\bf{p}},{\bf{k}} - {\bf{q}}/2 - {\bf{q}}^{'}) 
 - N({\bf{p}},{\bf{k}} + {\bf{q}}/2 - {\bf{q}}^{'}))
\langle n_{ {\bf{q}} }({\bf{k}}) \rangle
\]
\[
-(\langle n_{0}({\bf{k}} + {\bf{q}}/2) \rangle - 
\langle n_{0}({\bf{k}} - {\bf{q}}/2) \rangle)
\sum_{ {\bf{k}}^{'} \neq {\bf{k}} }\frac{ v_{ {\bf{k}}-{\bf{k}}^{'} } }{V}
\langle  \delta \mbox{   }n_{0}({\bf{p}}) n_{ {\bf{q}} }({\bf{k}}^{'}) \rangle
\]
\[
-(N({\bf{p}},{\bf{k}}+ {\bf{q}}/2) - N({\bf{p}},{\bf{k}}-{\bf{q}}/2) )
\sum_{ {\bf{k}}^{'} \neq {\bf{k}} }\frac{ v_{ {\bf{k}}-{\bf{k}}^{'} }}{V}
\langle  n_{ {\bf{q}} }({\bf{k}}^{'}) \rangle
\]
\begin{equation}
+ U_{ext}(-{\bf{q}},0)(N({\bf{p}},{\bf{k}}+ {\bf{q}}/2) -
 N({\bf{p}},{\bf{k}}- {\bf{q}}/2))
\end{equation}

\[
(\omega - {\tilde{\epsilon}}_{ {\bf{k}} - {\bf{q}}/2 }
 + {\tilde{\epsilon}}_{ {\bf{k}} + {\bf{q}}/2 })
\langle n_{ {\bf{q}} }({\bf{k}}) \rangle
 = 
 - \sum_{ {\bf{q}}^{'} \neq 0 }\frac{ v_{ {\bf{q}}^{'} } }{V}
(\langle \delta n_{0}( {\bf{k}}-{\bf{q}}/2 - {\bf{q}}^{'} )
n_{ {\bf{q}} }({\bf{k}}) \rangle
 - \langle \delta n_{0}( {\bf{k}}+{\bf{q}}/2 - {\bf{q}}^{'} )
n_{ {\bf{q}} }({\bf{k}})\rangle )
\]
\[
 -(\langle n_{0}({\bf{k}} + {\bf{q}}/2) \rangle
 - \langle n_{0}({\bf{k}} - {\bf{q}}/2) \rangle) 
\sum_{ {\bf{k}}^{'} \neq {\bf{k}} }
\frac{ v_{ {\bf{k}}-{\bf{k}}^{'} } }{V}
 \langle n_{ {\bf{q}} }({\bf{k}}^{'}) \rangle 
\]
\[
 -\sum_{ {\bf{k}}^{'} \neq {\bf{k}} }
\frac{ v_{ {\bf{k}}-{\bf{k}}^{'} } }{V}
(\langle \delta n_{0}({\bf{k}} + {\bf{q}}/2)
 n_{ {\bf{q}} }({\bf{k}}^{'}) \rangle
 - \langle \delta n_{0}({\bf{k}} - {\bf{q}}/2)
 n_{ {\bf{q}} }({\bf{k}}^{'}) \rangle )
\]
\begin{equation}
+ U_{ext}(-{\bf{q}},0)
 ( \langle n_{0}({\bf{k}} + {\bf{q}}/2) \rangle 
 - \langle n_{0}({\bf{k}} - {\bf{q}}/2) \rangle )
\end{equation}
 In order to simplify this further, let us assume that we are in the weakly
 nonideal regime. That is, it is legitimate to treat the momentum distribution
 as possessing a sharp Fermi surface and no other striking features. Let us now
 define,
\begin{equation}
 {\tilde{n}}_{ {\bf{q}} }({\vec{r}}) = \frac{1}{V}\sum_{ {\bf{k}} }
e^{-i{\bf{k}}.{\vec{r}} }n_{ {\bf{q}} }({\bf{k}})
\end{equation}
\[
(\omega + i\frac{ {\bf{q}}.\nabla }{m})
\langle \mbox{  }
\delta \mbox{  }n_{0}({\bf{p}})\mbox{  }
{\tilde{n}}_{ {\bf{q}} }({\vec{r}}) \rangle
 = -\frac{1}{V}\sum_{ {\bf{k}},{\bf{q}}^{'} \neq 0 }
\frac{ v_{ {\bf{q}}^{'} } }{V}
\int d^{3}r^{'}\mbox{  }v(r^{'})\mbox{  }
e^{i{\bf{k}}.({\vec{r}}^{'}-{\vec{r}})}
(N({\bf{p}},{\bf{k}} - {\bf{q}}/2 - {\bf{q}}^{'}) 
 - N({\bf{p}},{\bf{k}} + {\bf{q}}/2 - {\bf{q}}^{'}))
\mbox{   }\langle {\tilde{n}}_{ {\bf{q}} }({\vec{r}}^{'}) \rangle
\]
\[
-\frac{1}{V}\int d^{3}r^{'}\mbox{  }
\sum_{ {\bf{k}} }e^{i{\bf{k}}.({\vec{r}}^{'}-{\vec{r}})}
(\langle n_{0}({\bf{k}} + {\bf{q}}/2) \rangle - 
\langle n_{0}({\bf{k}} - {\bf{q}}/2) \rangle)v(r^{'})
\langle  \delta \mbox{   }n_{0}({\bf{p}}) n_{ {\bf{q}} }({\vec{r}}^{'}) \rangle
\]
\[
-\frac{1}{V}
\int d^{3}r^{'}\mbox{  }
\sum_{ {\bf{k}} }e^{i{\bf{k}}.({\vec{r}}^{'}-{\vec{r}})}
(N({\bf{p}},{\bf{k}}+ {\bf{q}}/2) - N({\bf{p}},{\bf{k}}-{\bf{q}}/2) )
v(r^{'}) \langle  n_{ {\bf{q}} }({\vec{r}}^{'}) \rangle
\]
\begin{equation}
+ U_{ext}(-{\bf{q}},0)\frac{1}{V}\sum_{ {\bf{k}} }e^{-i{\bf{k}}.{\vec{r}}}
(N({\bf{p}},{\bf{k}}+ {\bf{q}}/2) -
 N({\bf{p}},{\bf{k}}- {\bf{q}}/2))
\end{equation}

\[
(\omega + i\frac{ {\bf{q}}.\nabla }{m})
\langle {\tilde{n}}_{ {\bf{q}} }({\vec{r}}) \rangle
 = 
 -\frac{1}{V}
\sum_{ {\bf{k}},{\bf{q}}^{'} \neq 0 }\frac{ v_{ {\bf{q}}^{'} } }{V}
e^{i{\bf{k}}.({\vec{r}}^{'}-{\vec{r}})}
(\langle \delta n_{0}( {\bf{k}}-{\bf{q}}/2 - {\bf{q}}^{'} )
n_{ {\bf{q}} }({\bf{k}}) \rangle
 - \langle \delta n_{0}( {\bf{k}}+{\bf{q}}/2 - {\bf{q}}^{'} )
n_{ {\bf{q}} }({\vec{r}}^{'})\rangle )
\]
\[
 -\frac{1}{V}
\sum_{ {\bf{k}} }
\int d^{3}r^{'}\mbox{  }v(r^{'})
 e^{i{\bf{k}}.({\vec{r}}^{'}-{\vec{r}})}
(\langle n_{0}({\bf{k}} + {\bf{q}}/2) \rangle
 - \langle n_{0}({\bf{k}} - {\bf{q}}/2) \rangle) 
\langle n_{ {\bf{q}} }({\vec{r}}^{'}) \rangle 
\]
\[
 -\frac{1}{V}
\sum_{ {\bf{k}} } e^{i{\bf{k}}.({\vec{r}}^{'}-{\vec{r}})}
\int d^{3}r^{'}\mbox{  }v(r^{'})
(\langle \delta n_{0}({\bf{k}} + {\bf{q}}/2)
 n_{ {\bf{q}} }({\vec{r}}^{'}) \rangle
 - \langle \delta n_{0}({\bf{k}} - {\bf{q}}/2)
 n_{ {\bf{q}} }({\vec{r}}^{'}) \rangle )
\]
\begin{equation}
+ U_{ext}(-{\bf{q}},0)
\frac{1}{V}
\sum_{ {\bf{k}} } e^{-i{\bf{k}}.{\vec{r}}}
 ( \langle n_{0}({\bf{k}} + {\bf{q}}/2) \rangle 
 - \langle n_{0}({\bf{k}} - {\bf{q}}/2) \rangle )
\end{equation}
Let us now compute the following quantity,
\begin{equation}
f_{ {\bf{q}} }({\vec{R}}) = 
\frac{1}{V}\sum_{ {\bf{k}} }e^{i{\bf{k}}.({\vec{r}}^{'}-{\vec{r}})}
(\langle n_{0}({\bf{k}} + {\bf{q}}/2) \rangle - 
 \langle n_{0}({\bf{k}} - {\bf{q}}/2) \rangle) 
\end{equation}
where $ {\vec{R}} = {\vec{r}}^{'} - {\vec{r}} $.
If $ k_{f} $ is sufficiently large ( $ k_{f} >> q, k_{f} >> 1/R $) we may
 write,
\begin{equation}
f_{ {\bf{q}} }({\vec{R}}) \approx C_{0}{\bf{q}}.\nabla_{R}\delta^{3}(R)
\end{equation}
\begin{equation}
C_{0} = \frac{ (4\pi)^{2} }{ (2\pi)^{3} }
(\frac{i}{q})
\int^{\infty}_{0}
dR\mbox{  }\frac{2^{3}}{q^{2}R^{2}}
(sin(k_{f}R) - (k_{f}R)cos(k_{f}R))
(sin(\frac{qR}{2}) - (\frac{qR}{2})cos(\frac{qR}{2}))
\end{equation}
\[
(\omega + i\frac{ {\bf{q}}.\nabla }{m})
\langle 
\delta n_{0}({\bf{p}})\mbox{  }
{\tilde{n}}_{ {\bf{q}} }({\vec{r}}) \rangle
 = -\frac{1}{V}\sum_{ {\bf{k}},{\bf{q}}^{'} \neq 0 }
\frac{ v_{ {\bf{q}}^{'} } }{V}
\int d^{3}r^{'}\mbox{  }v(r^{'})\mbox{  }
e^{i{\bf{k}}.({\vec{r}}^{'}-{\vec{r}})}
(N({\bf{p}},{\bf{k}} - {\bf{q}}/2 - {\bf{q}}^{'}) 
 - N({\bf{p}},{\bf{k}} + {\bf{q}}/2 - {\bf{q}}^{'}))
\mbox{   }\langle {\tilde{n}}_{ {\bf{q}} }({\vec{r}}^{'}) \rangle
\]
\[
 + C_{0}{\bf{q}}.\nabla_{ {\vec{r}} }v(r)\langle \delta n_{0}({\bf{p}}) 
{\tilde{n}}_{ {\bf{q}} }({\vec{r}}) \rangle
\]
\[
-\frac{1}{V}\int d^{3}r^{'}\mbox{  }
\sum_{ {\bf{k}} }e^{i{\bf{k}}.({\vec{r}}^{'}-{\vec{r}})}
(N({\bf{p}},{\bf{k}}+ {\bf{q}}/2) - N({\bf{p}},{\bf{k}}-{\bf{q}}/2) )
v(r^{'}) \langle  n_{ {\bf{q}} }({\vec{r}}^{'}) \rangle
\]
\begin{equation}
+ U_{ext}(-{\bf{q}},0)\frac{1}{V}\sum_{ {\bf{k}} }e^{-i{\bf{k}}.{\vec{r}}}
(N({\bf{p}},{\bf{k}}+ {\bf{q}}/2) -
 N({\bf{p}},{\bf{k}}- {\bf{q}}/2))
\end{equation}

\[
(\omega + i\frac{ {\bf{q}}.\nabla }{m})
\langle {\tilde{n}}_{ {\bf{q}} }({\vec{r}}) \rangle
 = 
 -\frac{1}{V}
\sum_{ {\bf{k}},{\bf{q}}^{'} \neq 0 }\frac{ v_{ {\bf{q}}^{'} } }{V}
\int d^{3}r^{'} \mbox{   }v(r^{'})
e^{i{\bf{k}}.({\vec{r}}^{'}-{\vec{r}})}
(\langle \delta n_{0}( {\bf{k}}-{\bf{q}}/2 - {\bf{q}}^{'} )
n_{ {\bf{q}} }({\vec{r}}^{'}) \rangle
 - \langle \delta n_{0}( {\bf{k}}+{\bf{q}}/2 - {\bf{q}}^{'} )
n_{ {\bf{q}} }({\vec{r}}^{'})\rangle )
\]
\[
 + C_{0}{\bf{q}}.\nabla_{ {\vec{r}} }
v(r)\langle {\tilde{n}}_{ {\bf{q}} }({\vec{r}}) \rangle
\]
\[
 -\frac{1}{V}
\sum_{ {\bf{k}} } e^{i{\bf{k}}.({\vec{r}}^{'}-{\vec{r}})}
\int d^{3}r^{'}\mbox{  }v(r^{'})
(\langle \delta n_{0}({\bf{k}} + {\bf{q}}/2)
 n_{ {\bf{q}} }({\vec{r}}^{'}) \rangle
 - \langle \delta n_{0}({\bf{k}} - {\bf{q}}/2)
 n_{ {\bf{q}} }({\vec{r}}^{'}) \rangle )
\]
\begin{equation}
+ U_{ext}(-{\bf{q}},0)
\frac{1}{V}
\sum_{ {\bf{k}} } e^{-i{\bf{k}}.{\vec{r}}}
 ( \langle n_{0}({\bf{k}} + {\bf{q}}/2) \rangle 
 - \langle n_{0}({\bf{k}} - {\bf{q}}/2) \rangle )
\end{equation}
At this stage the author is unable to complete this calculation, but 
hopefully the reader appreciates the spirit of the discussion. 
 The author apologises in advance for this omission.

\section{Improved Theory of Single-Particle Properties Using Bosonization}

 In this section,
 we use the sea-boson method suitably generalised to accomodate
 fluctuations in the momentum distribution to compute
 single-particle properties.
 In the appendix we show how to derive the dielectric function
 using this method.
 The dielectric function may be evaluated by more conventional means as well
 \cite{Setlur2}. However, the single-particle properties evaluated using the
 sea-boson method is more accurate as one is able to treat the problem more
 systematically and the physical interpretation of the formulas is also
 more transparent. The really systematic approach would be to write down the
 sea-boson correspondence suitably generalised to accomodate fluctuations
 in the momentum distribution and evaluate the various propagators 
 like we did in our earlier work\cite{Setlur}. However, we shall adopt
 a simpler but hopefully the correct approach here. The idea is to simply
 borrow from our earlier work except that the zeros of the dielectric
 function are no longer the zeros of the RPA dielectric function but
 that of $ \epsilon_{eff} $ and we need to interpret the derivate
 of the dielectric funcion with respect to frequency that appears in
 these formulas as being the derivative of the full dielectric function.
 Therefore, let us write down the various
 formulas. The full hamiltonian in diagonalised form has the following 
 appearence,
\begin{equation}
H_{full} = \sum_{ {\bf{q}}, i }\omega_{i}({\bf{q}})d^{\dagger}_{i}({\bf{q}})
d_{i}({\bf{q}})
\label{DIAGFORM}
\end{equation}
\begin{equation}
{\bar{n}}_{ {\bf{k}} } =  n^{\beta}({\bf{k}}) \mbox{      }
F_{1}({\bf{k}})
 + (1-n^{\beta}({\bf{k}}) )\mbox{      } F_{2}({\bf{k}})
\end{equation}
where,
\begin{equation}
F_{1}({\bf{k}}) =  \frac{1}{1 + \frac{S_{B}({\bf{k}})}{1 + S_{A}({\bf{k}})}}
\end{equation}
\begin{equation}
F_{2}({\bf{k}}) =  \frac{1}{1 + \frac{1+S_{B}({\bf{k}})}{S_{A}({\bf{k}})}}
\end{equation}
\begin{equation}
S_{A}({\bf{k}}) = \sum_{ {\bf{q}}, i }
\frac{ {\bar{n}}_{ {\bf{k-q}} }  }
{(\omega_{i}(-{\bf{q}}) + {\bf{k.q}}/m - \epsilon_{ {\bf{q}} })^{2}}
g_{i}^{2}(-{\bf{q}})
\end{equation}
\begin{equation}
S_{B}({\bf{k}}) = 
  \sum_{ {\bf{q}},i }
\frac{ 1 - {\bar{n}}_{ {\bf{k+q}} }  }
{(\omega_{i}(-{\bf{q}}) + {\bf{k.q}}/m + \epsilon_{ {\bf{q}} })^{2}}
g_{i}^{2}(-{\bf{q}})
\end{equation}
\begin{equation}
g_{i}({\bf{q}}) = [\sum_{ {\bf{k}} }\frac{ {\bar{n}}_{ {\bf{k}}-{\bf{q}}/2 }
 - {\bar{n}}_{ {\bf{k}}+{\bf{q}}/2 } }
{ (\omega_{i}({\bf{q}}) - \frac{ {\bf{k.q}} }{m})^{2} }]^{-\frac{1}{2}}
\end{equation}
It may be observed that this quantity is just the frequency derivative
of the polarization. In other words,
\begin{equation}
g^{-2}_{i}({\bf{q}}) = \frac{V}{ v_{ {\bf{q}} } }
(\frac{ \partial }{ \partial \omega })_{ \omega = \omega_{i}({\bf{q}}) }
\epsilon^{P}_{g-RPA}({\bf{q}},\omega) 
\end{equation}
By analogy we may generalise this so that when fluctuations are introduced,
\begin{equation}
g^{-2}_{i}({\bf{q}}) = \frac{V}{v_{ {\bf{q}} } }
(\frac{ \partial }{ \partial \omega })_{\omega = \omega_{i}({\bf{q}}) }
\epsilon^{P}_{eff}({\bf{q}},\omega)
\end{equation}
 We shall employ this latter definition in our analysis.
 $ \omega_{i}({\bf{q}}) $ is the zero of the overall dielectric function,
\begin{equation}
\epsilon^{P}_{eff}({\bf{q}},\omega_{i}) = 0
\end{equation}
It is worthwhile simplifying the effective dielectric function.
\begin{equation}
\epsilon_{eff}({\bf{q}},\omega) = \epsilon_{g-RPA}({\bf{q}},\omega)
+ \frac{ ( \epsilon_{g-RPA}({\bf{q}},\omega) - 1 )
(\epsilon_{g-RPA}({\bf{q}},\omega) - \epsilon_{\beta}({\bf{q}},\omega)) }
 { \epsilon_{g-RPA}({\bf{q}},\omega) }
\label{EPSEFF1}
\end{equation}
The principal part of this is given by,
\begin{equation}
\epsilon^{P}_{eff}({\bf{q}},\omega) = \epsilon^{P}_{g-RPA}({\bf{q}},\omega)
+ \frac{ ( \epsilon^{P}_{g-RPA}({\bf{q}},\omega) - 1 )
(\epsilon^{P}_{g-RPA}({\bf{q}},\omega)
 - \epsilon^{P}_{\beta}({\bf{q}},\omega)) }
 { \epsilon^{P}_{g-RPA}({\bf{q}},\omega) }
\end{equation}
where,
\begin{equation}
\epsilon_{g-RPA}({\bf{q}},\omega)  = 1 + \frac{ v_{ {\bf{q}} } }{V}
\sum_{ {\bf{k}} }\frac{ {\bar{n}}_{ {\bf{k}}+{\bf{q}}/2 }
 - {\bar{n}}_{ {\bf{k}}-{\bf{q}}/2 } }
{\omega - \frac{ {\bf{k.q}} }{m}}
\end{equation}
\begin{equation}
\epsilon_{\beta}({\bf{q}},\omega)  = 1 + \frac{ v_{ {\bf{q}} } }{V}
\sum_{ {\bf{k}} }\frac{ n^{\beta}_{ {\bf{k}}+{\bf{q}}/2 }
 - n^{\beta}_{ {\bf{k}}-{\bf{q}}/2 } }
{\omega - \frac{ {\bf{k.q}} }{m}}
\end{equation}
and the principal parts of these functions are just the real parts.
\begin{equation}
\epsilon^{P}_{g-RPA}({\bf{q}},\omega)  = 1 + \frac{ v_{ {\bf{q}} } }{V}
\sum_{ {\bf{k}} }\mbox{       }
P\mbox{     }\frac{ {\bar{n}}_{ {\bf{k}}+{\bf{q}}/2 }
 - {\bar{n}}_{ {\bf{k}}-{\bf{q}}/2 } }
{\omega - \frac{ {\bf{k.q}} }{m}}
\end{equation}

\subsection{Just How Many Zeros Does the Dielectric Function Really Have ?}

 In this section, we address what is perhaps the most vexing problem in
 this approach. What is the size of the set to which $ i $ in 
 Eq.(~\ref{DIAGFORM}) belongs ? If one counts only the collective mode
 which is what a naive approach would lead us to do, then we are ignoring
 a large (infinite !) number of particle-hole modes. And yet, it seems
 that the particle-hole modes don't come about naturally and must be forced
 into the formalism\cite{Setlur}. It is really important to address this
 issue since ignoring it would mean that the excitation spectrum
 of the homogeneous electron gas possesses a gap when in fact it should not.
 In other words, if one counts only the collective mode(plasmon) one arrives at
 the unavoidable conclusion that the excitation spectrum has a finite
 gap(equal to the plasmon energy). How then should one introduce the
 particle-hole mode ? In this section, we show how to introduce both the
 particle-hole mode and collective mode in a unified manner. In our previous
 work\cite{Setlur}
 we suggested that the zeros of the RPA dielectric function should be
 interpreted as the maxima of the dynamical structure factor. This is quite
 an unusual and drastic departure from the notion of a root of a function.
 Being physicists we accept it as it has a physical interpretation. 
 Let us examine the diagonalised form of the full hamiltonian\cite{Setlur}:
\begin{equation}
H_{full} = \sum_{ {\bf{q}} , i }\omega_{i}({\bf{q}})
d^{\dagger}_{i}({\bf{q}})d_{i}({\bf{q}})
\end{equation}
 For small $ {\bf{q}} $, we expect this sum 
 to involve just the collective mode. In other  words,
\begin{equation}
H_{full} = \sum_{ {\bf{q}}=small }\omega_{c}({\bf{q}})
d^{\dagger}_{c}({\bf{q}})d_{c}({\bf{q}})
\end{equation}
This may be achieved by employing the following device,
\begin{equation}
H_{full} = \sum_{ {\bf{q}} }\int_{0}^{\infty}\mbox{     }
d\omega\mbox{       }\omega\mbox{         }W({\bf{q}},\omega)
\mbox{         }d^{\dagger}_{\omega}({\bf{q}})d_{\omega}({\bf{q}})
\end{equation}
Now these operators, the dressed sea-bosons in the new language obey a
 different kind of commutation rule:
\begin{equation}
[d_{\omega}({\bf{q}}), d_{\omega^{'}}({\bf{q}}^{'})] = 0
\end{equation}
\begin{equation}
[d_{\omega}({\bf{q}}), d^{\dagger}_{\omega^{'}}({\bf{q}}^{'})] = 
\delta_{\omega,\omega^{'}}\delta_{ {\bf{q}}, {\bf{q}}^{'} }
\end{equation}
 where $ \delta_{\omega,\omega^{'}} $ is a Kronecker delta function.
 Also the object $ W({\bf{q}},\omega) $ is a weight suitably chosen so that the
 the sum reduces to just the one over the collective mode in the small
 $ {\bf{q}} $ limit. It is also clear that $ W({\bf{q}},\omega) $ has to
 have dimensions of inverse energy.
 Let us therefore postulate a form,
\begin{equation}
W({\bf{q}},\omega) = -c_{0}\mbox{      }Im(\frac{1}{\epsilon({\bf{q}},\omega)})
\end{equation}
where $ c_{0} $ is a suitable constant. We know from textbooks
\cite{Mahan} that(in 3D),
\begin{equation}
Limit_{ {\bf{q}} \rightarrow 0 }
W({\bf{q}},\omega) = -c_{0}\mbox{      }
Limit_{ {\bf{q}} \rightarrow 0 }\mbox{      }
Im(\frac{1}{\epsilon({\bf{q}},\omega)})
 = \frac{ c_{0}\pi\omega_{p} }{2}\delta(\omega-\omega_{p})
\end{equation}
In order that we reproduce the right collective mode we must choose,
\begin{equation}
c_{0} = \frac{2}{\pi\omega_{p}}
\end{equation}
 The choices for $ W({\bf{q}},\omega) $ in other dimensions are worked out
 in the appendix. Now we have in our hands a 
 convenient way of labeling excited states of our system. The eigenstates of
 the system have energy eigenvalues given by,
\begin{equation}
\Omega_{\omega}({\bf{q}}) = \omega\mbox{      }W({\bf{q}},\omega)
\mbox{    }\Delta\omega
\end{equation}
 The spacing $ \Delta\omega $ is the smallest possible spacing between the
 energies, which is arbitrarily small. Thus we can see that, in most cases
 it does not cost a finite energy to excite the system. This corresponds to
 the particle-hole mode. On the other hand, if the weight $ W({\bf{q}},\omega) $
 is singular as it happens in the vicinity of $ \omega = \omega_{p} $and
 $ {\bf{q}} = 0 $, we have a finite gap.
\begin{equation}
\Omega_{\omega \approx \omega_{p}}({\bf{q}} \approx 0)
 = \omega_{p}\mbox{      }\delta(\omega-\omega_{p})
\mbox{    }\Delta\omega
\end{equation}
 Now if $ Lim_{\omega \rightarrow \omega_{p} } \delta(\omega-\omega_{p})\mbox{    }\Delta\omega = 1$
 (this defines $ \Delta\omega $ if you like), then we see the emergence of
 the collective mode. In gapped systems we expect only the collective mode
 and no particle hole mode. This is possible only if for any $ \omega \neq 0 $,
\begin{equation}
W({\bf{q}},\omega) = \infty
\end{equation}
 For the homogeneous Fermi system,
 this situation corresponds to Wigner crystallisation. In a Wigner crystal,
 in order to create excited states, you need to create phonons which require
 a nonzero-amount of energy each time.

 Thus, all energies are allowed but each comes with a "weight" corresponding
 to how strong the structure factor is at that energy. It may be seen that
 in 3-dimensions this would mean that for small $ {\bf{q}} $ the 
 sum over $ i $ is just the collective mode, but for larger $ {\bf{q}} $,
 we start summing the particle hole modes as well. The sums in the
 objects $ S_{A} $ and $ S_{B} $ have to be reinterpreted as well. 
 It is rather unfortunate that simple and straightforward interpretations
 are not possible, and one is forced to take recourse to devious means
 such as the one we shall now describe. Let us postulate, 
\begin{equation}
S_{A}({\bf{k}}) = \sum_{ {\bf{q}} }\int_{0}^{\infty}
\mbox{  }d\omega\mbox{   }{\tilde{W}}({\bf{q}},\omega)\mbox{  }
\frac{ {\bar{n}}_{ {\bf{k+q}} }  }
{(\omega - {\bf{k.q}}/m - \epsilon_{ {\bf{q}} })^{2}}
g_{\omega}^{2}({\bf{q}})
\end{equation}
\begin{equation}
S_{B}({\bf{k}}) = 
  \sum_{ {\bf{q}} }
\int_{0}^{\infty}
\mbox{  }d\omega\mbox{   }{\tilde{W}}({\bf{q}},\omega)\mbox{  }
\frac{ 1 - {\bar{n}}_{ {\bf{k-q}} }  }
{(\omega - {\bf{k.q}}/m + \epsilon_{ {\bf{q}} })^{2}}
g_{\omega}^{2}({\bf{q}})
\end{equation}
\begin{equation}
g_{\omega}^{2}({\bf{q}}) = \frac{v_{ {\bf{q}} }}{V}\frac{1}
{\frac{\partial}{\partial \omega}\epsilon^{P}_{eff}({\bf{q}},\omega)}
\end{equation}
and,
\begin{equation}
{\tilde{W}}({\bf{q}},\omega)\mbox{  } = W({\bf{q}},\omega)/
\int^{\infty}_{0}\mbox{     }d\omega\mbox{     }W({\bf{q}},\omega)
\end{equation}
 This seemingly adhoc ansatz becomes more plausible when one realises that
in the small $ {\bf{q}} $ limit it has a familiar form corresponding to
the collective mode.
\begin{equation}
S_{A}({\bf{k}}) \approx \sum_{ {\bf{q}}=small  }
\frac{ {\bar{n}}_{ {\bf{k+q}} }  }
{(\omega_{c}({\bf{q}}) - {\bf{k.q}}/m - \epsilon_{ {\bf{q}} })^{2}}
g_{c}^{2}({\bf{q}})
\end{equation}
\begin{equation}
S_{B}({\bf{k}}) \approx
  \sum_{ {\bf{q}}=small }
\frac{ 1 - {\bar{n}}_{ {\bf{k-q}} }  }
{(\omega_{c}({\bf{q}}) - {\bf{k.q}}/m + \epsilon_{ {\bf{q}} })^{2}}
g_{c}^{2}({\bf{q}})
\end{equation}
 Let us now move on to the full propagator.

\subsection{The Full Propagator }

 Let us borrow from our earlier work\cite{Setlur} and write down the final
 formulas,
\begin{equation}
\langle \psi^{\dagger}({\bf{x}},t)\psi({\bf{x}}^{'},t^{'})\rangle
 = |{\mathcal{R}}_{0}|^{2}|{\mathcal{Z}}_{0}|^{4}
e^{\sum_{ {\bf{k}}, {\bf{q}}, i }
U^{*i}_{ {\bf{k}}, {\bf{q}} }({\bf{x}})
U^{i}_{ {\bf{k}}, {\bf{q}} }({\bf{x}}^{'})
e^{i\mbox{ }{\tilde{\omega}}_{i}({\bf{q}})(t^{'}-t)}}
e^{-\sum_{ {\bf{k}}, {\bf{q}} }
g^{*}_{ {\bf{k}}, {\bf{q}} }({\bf{x}})
g_{ {\bf{k}}, {\bf{q}} }({\bf{x}}^{'})
e^{i\mbox{ }\omega_{ {\bf{k}} }({\bf{q}})(t^{'}-t)}}
\langle \psi^{\dagger}({\bf{x}},t)\psi({\bf{x}}^{'},t^{'})\rangle_{free}
\end{equation}
\begin{equation}
\langle \psi({\bf{x}}^{'},t^{'})\psi^{\dagger}({\bf{x}},t)\rangle
 = |{\mathcal{R}}_{0}|^{2}|{\mathcal{Z}}_{0}|^{4}
e^{\sum_{ {\bf{k}}, {\bf{q}}, i }
U^{*i}_{ {\bf{k}}, {\bf{q}} }({\bf{x}}^{'})
U^{i}_{ {\bf{k}}, {\bf{q}} }({\bf{x}})
e^{i\mbox{ }{\tilde{\omega}}_{i}({\bf{q}})(t-t^{'})}}
e^{-\sum_{ {\bf{k}}, {\bf{q}} }
f^{*}_{ {\bf{k}}, {\bf{q}} }({\bf{x}})
f_{ {\bf{k}}, {\bf{q}} }({\bf{x}}^{'})
e^{i\mbox{ }\omega_{ {\bf{k}} }({\bf{q}})(t-t^{'})}}
\langle \psi({\bf{x}}^{'},t^{'})\psi^{\dagger}({\bf{x}},t)\rangle_{free}
\end{equation}
 In the above formula, the index $ i $ runs over both the collective
 mode as well as the particle-hole modes($ i =  c, {\bf{k}}_{i} $). As we have
 pointed out earlier, this must be reinterpreted to mean a weighted sum
 with the dynamical structure factor suitably normalised appearing as the
 weight.
\begin{equation}
g_{ {\bf{k}}, {\bf{q}} }({\bf{x}}) = -e^{-i\mbox{ }{\bf{q.x}} }
(\frac{1}{2\mbox{ }N\mbox{ }\epsilon_{ {\bf{q}} } })
\Lambda^{0}_{ {\bf{k}} }(-{\bf{q}})\omega_{ {\bf{k}} }({\bf{q}})
 + i\mbox{ }U_{ {\bf{q}} }({\bf{x}})\Lambda^{0}_{ {\bf{k}} }(-{\bf{q}})
= -f^{*}_{ {\bf{k}}, {\bf{q}} }({\bf{x}})
\end{equation}
Set $ U_{ {\bf{q}} }({\bf{x}}) = e^{-i\mbox{ }{\bf{q.x}}}U_{0}({\bf{q}}) $ 
\begin{equation}
U_{0}({\bf{q}}) = \frac{1}{N}
(\frac{ \theta(k_{f} - |{\bf{q}}|) - w_{1}({\bf{q}}) }{ w_{2}({\bf{q}}) })
^{\frac{1}{2}}
\end{equation}
\begin{equation}
w_{1}({\bf{q}}) = (\frac{1}{4\mbox{ }N\mbox{ }\epsilon^{2}_{ {\bf{q}} }})
\sum_{ {\bf{k}} }(\frac{ {\bf{k.q}} }{m})^{2}
(\Lambda^{0}_{ {\bf{k}} }(-{\bf{q}}))^{2}
\end{equation}
\begin{equation}
w_{2}({\bf{q}}) = (\frac{1}{N})\sum_{ {\bf{k}} }
(\Lambda^{0}_{ {\bf{k}} }(-{\bf{q}}))^{2}
\end{equation}
 Here $ \Lambda^{0}_{ {\bf{k}} }(-{\bf{q}}) $ is the $ \Lambda $
 with a noninteracting momentum distribution.
The interacting coefficients are given as,
\begin{equation}
U^{i}_{ {\bf{k}}, {\bf{q}} }({\bf{x}}) =
f^{*}_{ {\bf{k}}, {\bf{q}} }({\bf{x}})
[a_{ {\bf{k}} }({\bf{q}}), b^{\dagger}_{i}({\bf{q}})]
+ f_{ {\bf{k}}, -{\bf{q}} }({\bf{x}})
[a_{ {\bf{k}} }(-{\bf{q}}), b_{i}({\bf{q}})]
\end{equation}
\[
{\mathcal{R}}_{0} =
exp(-\sum_{ {\bf{k}}, {\bf{q}}, i }
f^{*}_{ {\bf{k}}, {\bf{q}} }({\bf{x}})f_{ {\bf{k}}, {\bf{q}} }({\bf{x}})
[b_{i}({\bf{q}}), a^{\dagger}_{ {\bf{k}} }({\bf{q}})]
[a_{ {\bf{k}} }({\bf{q}}), b^{\dagger}_{i}({\bf{q}})])
\]
\[
\times
exp(-\frac{1}{2}\sum_{ {\bf{k}}, {\bf{q}}, i }
 f^{*}_{ {\bf{k}}, {\bf{q}} }({\bf{x}})f^{*}_{ {\bf{k}}, -{\bf{q}} }({\bf{x}})
[a_{ {\bf{k}} }(-{\bf{q}}), b_{i}({\bf{q}})]
[a_{ {\bf{k}} }({\bf{q}}), b^{\dagger}_{i}({\bf{q}})])
\]
\begin{equation}
\times
exp(-\frac{1}{2}\sum_{ {\bf{k}}, {\bf{q}}, i }
 f_{ {\bf{k}}, {\bf{q}} }({\bf{x}})f_{ {\bf{k}}, -{\bf{q}} }({\bf{x}})
[a_{ {\bf{k}} }(-{\bf{q}}), b_{i}({\bf{q}})]
[a_{ {\bf{k}} }({\bf{q}}), b^{\dagger}_{i}({\bf{q}})])
\end{equation}
\begin{equation}
{\mathcal{Z}}_{0} = e^{i\mbox{ }\sum_{ {\bf{k}}, {\bf{q}} \neq 0 }
U_{0}({\bf{q}})(\frac{1}{2 \mbox{ }N\epsilon_{ {\bf{q}} }})
(\Lambda^{0}_{ {\bf{k}} }(-{\bf{q}}))^{2}\omega_{ {\bf{k}} }({\bf{q}}) }
 e^{\frac{1}{2}
\mbox{ }\sum_{ {\bf{k}}, {\bf{q}} \neq 0 }
(\frac{1}{2 \mbox{ }N\epsilon_{ {\bf{q}} }})^{2}
(\Lambda^{0}_{ {\bf{k}} }(-{\bf{q}}))^{2}(\omega_{ {\bf{k}} }({\bf{q}}))^{2} }
e^{\frac{1}{2}
\mbox{ }\sum_{ {\bf{k}}, {\bf{q}} \neq 0 }
(U_{0}({\bf{q}}))^{2}
(\Lambda^{0}_{ {\bf{k}} }(-{\bf{q}}))^{2} }
\end{equation}
The commutators are given as before,
\begin{equation}
[b_{i}({\bf{q}}), a^{\dagger}_{ {\bf{k}} }({\bf{q}})]
 = (\frac{ \Lambda_{ {\bf{k}} }(-{\bf{q}}) }
{ {\tilde{\omega}}_{i}({\bf{q}}) - \frac{ {\bf{k.q}} }{m} })
g_{i}({\bf{q}}) = [a_{ {\bf{k}} }({\bf{q}}), b^{\dagger}_{i}({\bf{q}})]
\end{equation}
\begin{equation}
[b_{i}({\bf{q}}), a_{ {\bf{k}} }(-{\bf{q}})]
 = -(\frac{ \Lambda_{ {\bf{k}} }({\bf{q}}) }
{ {\tilde{\omega}}_{i}({\bf{q}}) - \frac{ {\bf{k.q}} }{m} })
g_{i}({\bf{q}})
\end{equation}
\begin{equation}
[a_{ {\bf{k}} }({\bf{q}}), b_{i}(-{\bf{q}})]
 = (\frac{ \Lambda_{ {\bf{k}} }(-{\bf{q}}) }
{ {\tilde{\omega}}_{i}(-{\bf{q}}) + \frac{ {\bf{k.q}} }{m} })
g_{i}(-{\bf{q}})
\end{equation}
\begin{equation}
g^{2}_{i}({\bf{q}}) =  \frac{v_{ {\bf{q}} }}{V}\frac{1}
{\frac{\partial}{\partial \omega}|_{\omega = \omega_{i} }
\epsilon^{P}_{eff}({\bf{q}},\omega)}
\end{equation}

\subsection{Exchange Effects within RPA}

 It may be possible to employ a myriad of other similar approaches each
 differing from the other in how the notion of RPA is implemented. To give
 an example, let us rewrite the full hamiltonian differently.
\[
H = \sum_{ {\bf{k}} }\epsilon_{ {\bf{k}} }c^{\dagger}_{ {\bf{k}} }
c_{ {\bf{k}} }
 - \sum_{ {\bf{q}} \neq 0 }\frac{v({\bf{q}})}{2V}
\sum_{ {\bf{k}} }
n_{ {\bf{k}} + {\bf{q}}/2 }n_{ {\bf{k}} - {\bf{q}}/2 }
\]
\begin{equation}
+ \sum_{ {\bf{q}} \neq 0 }\frac{v({\bf{q}})}{2V}
\sum_{ {\bf{k}} \neq {\bf{k}}^{'} }n_{ {\bf{q}} }({\bf{k}})
n_{ -{\bf{q}} }({\bf{k}}^{'})
\end{equation}
 Here the exchange term has been singled out for special treatment. We could
 rewrite the last term differently,
\[
H = \sum_{ {\bf{k}} }\epsilon_{ {\bf{k}} }c^{\dagger}_{ {\bf{k}} }
c_{ {\bf{k}} }
 - \sum_{ {\bf{q}} \neq 0 }\frac{v({\bf{q}})}{2V}
\sum_{ {\bf{k}} }
n_{ {\bf{k}} + {\bf{q}}/2 }n_{ {\bf{k}} - {\bf{q}}/2 }
\]
\begin{equation}
- \sum_{ {\bf{q}} \neq 0 }\frac{v({\bf{q}})}{2V}
\sum_{ {\bf{k}} \neq {\bf{k}}^{'} }n_{ {\bf{k}}-{\bf{k}}^{'} }
({\bf{k}}/2 + {\bf{k}}^{'}/2 + {\bf{q}}/2 )
n_{ {\bf{k}}^{'}-{\bf{k}} }({\bf{k}}/2 + {\bf{k}}^{'}/2 - {\bf{q}}/2)
\end{equation}
 This may be interpreted as an exchange term. The negative sign suggests 
 precisely this.
 These two hamiltonians lead to different answers when the RPA is carried out
 on them. In fact, even the exchange term involving just the
 number operators can be treated in two different ways.
 One way is to retain it as it is, the other is to rewrite it as,
\begin{equation}
H_{ex} = \sum_{ {\bf{k}} }{\tilde{\epsilon}}_{ {\bf{k}} }
c^{\dagger}_{ {\bf{k}} }
c_{ {\bf{k}} }
 - \sum_{ {\bf{q}} \neq 0 }\frac{v({\bf{q}})}{2V}
\sum_{ {\bf{k}} }
\delta\mbox{    }n_{ {\bf{k}} + {\bf{q}}/2 }
\mbox{       }\delta\mbox{   }n_{ {\bf{k}} - {\bf{q}}/2 }
\end{equation}
 The effective dispersion includes the exchange energy,
\begin{equation}
{\tilde{\epsilon}}_{ {\bf{k}} } = \epsilon_{ {\bf{k}} }
 - \sum_{ {\bf{q}} \neq 0 } \frac{ v({\bf{q}}) }{V}
\langle n_{ {\bf{k-q}} } \rangle
\end{equation}
 and
 $ \delta\mbox{   }n_{ {\bf{k}} } = n_{ {\bf{k}} } - {\bar{n}}_{ {\bf{k}} }  $
 Again when the RPA is carried out on them they yield different answers.
 One has to examine this issue more closely. In particular the following
 questions spring to mind. Should we use one or the other or a combination
 of the two ? The answer may be given by appealing to our bosonization
 scheme.
For more details, the reader is refered to our earlier works\cite{Setlur}
,\cite{PREPRINT2}. Just to make this article self-contained, it is desirable
 to fix some notation.
\begin{equation}
n_{0}({\bf{k}}) = c^{\dagger}_{ {\bf{k}} }c_{ {\bf{k}} }
\end{equation}
the $ a_{ {\bf{k}} }({\bf{q}}) $ is a sea-boson corresponding to the fermions
 $ c_{ {\bf{k}} } $. The coefficients
 $ \Lambda_{ {\bf{k}} }({\bf{q}}) = \sqrt{{\bar{n}}_{ {\bf{k}} + {\bf{q}}/2 }(1 - {\bar{n}}_{ {\bf{k}} - {\bf{q}}/2 } ) } $
 are c-numbers for now.
\begin{equation}
c^{\dagger}_{ {\bf{k}} + {\bf{q}}/2 } c_{ {\bf{k}}-{\bf{q}}/2 }
 = \Lambda_{ {\bf{k}} }({\bf{q}}) a_{ {\bf{k}} }(-{\bf{q}})
 + a^{\dagger}_{ {\bf{k}} }({\bf{q}}) \Lambda_{ {\bf{k}} }(-{\bf{q}})
\end{equation}
Let us now try and figure out how to decompose products of four fermions,
(let us further assume that none of the four indices are equal to
 any other index)($ {\bf{q}} \neq 0 $ and $ {\bf{k}} \neq {\bf{k}}^{'} $)
\begin{equation}
c^{\dagger}_{ {\bf{k}} + {\bf{q}}/2 } c_{ {\bf{k}}-{\bf{q}}/2 }
c^{\dagger}_{ {\bf{k}}^{'} - {\bf{q}}/2 } c_{ {\bf{k}}^{'}+{\bf{q}}/2 }
 = [\Lambda_{ {\bf{k}} }({\bf{q}}) a_{ {\bf{k}} }(-{\bf{q}})
 + a^{\dagger}_{ {\bf{k}} }({\bf{q}}) \Lambda_{ {\bf{k}} }(-{\bf{q}})]
 [\Lambda_{ {\bf{k}}^{'} }(-{\bf{q}}) a_{ {\bf{k}}^{'} }({\bf{q}})
 + a^{\dagger}_{ {\bf{k}}^{'} }(-{\bf{q}}) \Lambda_{ {\bf{k}}^{'} }({\bf{q}})]
\label{ANS1}
\end{equation}
 the above equality is within RPA. It may be noted that a slightly different
 regrouping yields a totally different formula,
\[
c^{\dagger}_{ {\bf{k}} + {\bf{q}}/2 } c_{ {\bf{k}}-{\bf{q}}/2 }
c^{\dagger}_{ {\bf{k}}^{'} - {\bf{q}}/2 } c_{ {\bf{k}}^{'}+{\bf{q}}/2 }
 = -c^{\dagger}_{ {\bf{k}} + {\bf{q}}/2 }
 c_{ {\bf{k}}^{'}+{\bf{q}}/2 }
c^{\dagger}_{ {\bf{k}}^{'} - {\bf{q}}/2 }
 c_{ {\bf{k}}-{\bf{q}}/2 }
\]
\[
=
 -[\Lambda_{ \frac{ {\bf{k}} }{2} + \frac{ {\bf{k}}^{'} }{2}
+ {\bf{q}}/2   }({\bf{k}}-{\bf{k}}^{'}) a_{\frac{ {\bf{k}} }{2}
 + \frac{ {\bf{k}}^{'} }{2}
+ \frac{ {\bf{q}} }{2} }({\bf{k}}^{'}-{\bf{k}})
 + a^{\dagger}_{ \frac{ {\bf{k}} }{2} + \frac{ {\bf{k}}^{'} }{2}
+ \frac{ {\bf{q}} }{2}  }({\bf{k}}-{\bf{k}}^{'})
 \Lambda_{ \frac{ {\bf{k}} }{2} + \frac{ {\bf{k}}^{'} }{2}
+ \frac{ {\bf{q}} }{2}   }({\bf{k}}^{'}-{\bf{k}})]
\]
\begin{equation}
 \times [\Lambda_{ \frac{ {\bf{k}} }{2} + \frac{ {\bf{k}}^{'} }{2}
- \frac{ {\bf{q}} }{2}   }({\bf{k}}^{'}-{\bf{k}}) a_{\frac{ {\bf{k}} }{2}
 + \frac{ {\bf{k}}^{'} }{2}
- \frac{ {\bf{q}} }{2} }({\bf{k}}-{\bf{k}}^{'})
 + a^{\dagger}_{ \frac{ {\bf{k}} }{2} + \frac{ {\bf{k}}^{'} }{2}
- \frac{ {\bf{q}} }{2}  }({\bf{k}}^{'}-{\bf{k}})
 \Lambda_{ \frac{ {\bf{k}} }{2} + \frac{ {\bf{k}}^{'} }{2}
- \frac{ {\bf{q}} }{2}   }({\bf{k}}-{\bf{k}}^{'})]
\label{ANS2}
\end{equation}
Let us now examine the commutator obtained using Fermi algebra,
\[
[c^{\dagger}_{ {\bf{k}} -{\bf{q}}/2 }c_{  {\bf{k}} + {\bf{q}}/2 },
c^{\dagger}_{ {\bf{k}} + {\bf{q}}/2 } c_{ {\bf{k}}-{\bf{q}}/2 }
c^{\dagger}_{ {\bf{k}}^{'} - {\bf{q}}/2 } c_{ {\bf{k}}^{'}+{\bf{q}}/2 }]
 = [n_{0}({\bf{k}} -{\bf{q}}/2)-n_{0}({\bf{k}} +{\bf{q}}/2)]
c^{\dagger}_{ {\bf{k}}^{'} - {\bf{q}}/2 } c_{ {\bf{k}}^{'}+{\bf{q}}/2 }
\]
\begin{equation}
+
[ \delta_{ {\bf{k}}^{'} - {\bf{q}}/2, {\bf{k}} + {\bf{q}}/2 }
(1-n_{0}( {\bf{k}} - {\bf{q}}/2 ))
c^{\dagger}_{ {\bf{k}} + {\bf{q}}/2 }
c_{ {\bf{k}}^{'} + {\bf{q}}/2 }
 + \delta_{  {\bf{k}} - {\bf{q}}/2, {\bf{k}}^{'} + {\bf{q}}/2 }
 n_{0}( {\bf{k}} + {\bf{q}}/2 )
c^{\dagger}_{ {\bf{k}}^{'} - {\bf{q}}/2 } c_{ {\bf{k}} - {\bf{q}}/2 } ]
\label{COMM}
\end{equation}
 It is clear that if one uses the ansatz in Eq.(~\ref{ANS1}) then one obtains
 only the first term in Eq.(~\ref{COMM}). In order to obtain the second
 term it is necessary to go beyond this and try and include some portion of
 Eq.(~\ref{ANS2}). Let us first try to use the ansatz
 in Eq.(~\ref{ANS2}) while completely ignoring the ansatz of Eq.(~\ref{ANS1})
 In this case the interaction term takes a different form,
\begin{equation}
H_{I} = -\sum_{ {\bf{q}} \neq 0 }
\sum_{ {\bf{k}} \neq {\bf{k}}^{'} }
\frac{v_{ {\bf{k}}-{\bf{k}}^{'} } }{2V}
[\Lambda_{ {\bf{k}} }({\bf{q}})a_{ {\bf{k}} }(-{\bf{q}})
 + a^{\dagger}_{ {\bf{k}} }({\bf{q}})
\Lambda_{ {\bf{k}} }(-{\bf{q}})]
[\Lambda_{ {\bf{k}}^{'} }(-{\bf{q}})a_{ {\bf{k}}^{'} }({\bf{q}})
 + a^{\dagger}_{ {\bf{k}}^{'} }(-{\bf{q}})
\Lambda_{ {\bf{k}}^{'} }({\bf{q}})]
\end{equation}
 Here two important changes have occured. First, the interaction
 $ v_{ {\bf{q}} } $ is replaced by $ v_{ {\bf{k}} - {\bf{k}}^{'} } $
 and there is a change in sign. The physical meaning of these mysterious
 changes will be defered until later when we have investigated the details.
 Let us now write down the full hamiltonian,
\begin{equation}
H = \sum_{ {\bf{k}}, {\bf{q}} }{\tilde{\omega}}_{ {\bf{k}} }({\bf{q}})
a^{\dagger}_{ {\bf{k}} }({\bf{q}})a_{ {\bf{k}} }({\bf{q}})
-\sum_{ {\bf{q}} \neq 0 }
\sum_{ {\bf{k}} \neq {\bf{k}}^{'} }
\frac{v_{ {\bf{k}}-{\bf{k}}^{'} } }{2V}
[\Lambda_{ {\bf{k}} }({\bf{q}})a_{ {\bf{k}} }(-{\bf{q}})
 + a^{\dagger}_{ {\bf{k}} }({\bf{q}})
\Lambda_{ {\bf{k}} }(-{\bf{q}})]
[\Lambda_{ {\bf{k}}^{'} }(-{\bf{q}})a_{ {\bf{k}}^{'} }({\bf{q}})
 + a^{\dagger}_{ {\bf{k}}^{'} }(-{\bf{q}})
\Lambda_{ {\bf{k}}^{'} }({\bf{q}})]
\end{equation}
Here,
\begin{equation}
{\tilde{\omega}}_{ {\bf{k}} }({\bf{q}})
 = {\tilde{\epsilon}}_{ {\bf{k}} + {\bf{q}}/2 }
 - {\tilde{\epsilon}}_{ {\bf{k}} - {\bf{q}}/2 }
\end{equation}
\begin{equation}
{\tilde{\epsilon}}_{ {\bf{k}} } = \epsilon_{ {\bf{k}} }
 - \sum_{ {\bf{q}} \neq 0 }\frac{ v_{ {\bf{q}} } }{V}
{\bar{n}}_{ {\bf{k-q}} }
\end{equation}
Let us diagonalise this via a Bogoliubov transformation.
 To this end one assumes that the diagonalised
 form has the following apperance,
\begin{equation}
H = \sum_{ i, {\bf{q}} }\omega_{i}({\bf{q}})d^{\dagger}_{i}({\bf{q}})
d_{i}({\bf{q}})
\end{equation}
\[
\omega_{i}({\bf{q}})d_{i}({\bf{q}})
 = \sum_{ {\bf{k}}, {\bf{q}} }{\tilde{\omega}}_{ {\bf{k}} }({\bf{q}})
[d_{i}({\bf{q}}),a^{\dagger}_{ {\bf{k}} }({\bf{q}})]a_{ {\bf{k}} }({\bf{q}})
+  \sum_{ {\bf{k}}, {\bf{q}} }{\tilde{\omega}}_{ {\bf{k}} }(-{\bf{q}})
[d_{i}({\bf{q}}),a_{ {\bf{k}} }(-{\bf{q}})]
a^{\dagger}_{ {\bf{k}} }(-{\bf{q}})
\]
\begin{equation}
-\sum_{ {\bf{q}} \neq 0 }
\sum_{ {\bf{k}} \neq {\bf{k}}^{'} }
\frac{v_{ {\bf{k}}-{\bf{k}}^{'} } }{V}
(\Lambda_{ {\bf{k}}^{'} }({\bf{q}})
[d_{i}({\bf{q}}),a_{ {\bf{k}}^{'} }(-{\bf{q}})]
 + [d_{i}({\bf{q}}),a^{\dagger}_{ {\bf{k}}^{'} }({\bf{q}})]
\Lambda_{ {\bf{k}}^{'} }(-{\bf{q}}))
(\Lambda_{ {\bf{k}} }(-{\bf{q}})a_{ {\bf{k}} }({\bf{q}})
 + a^{\dagger}_{ {\bf{k}} }(-{\bf{q}})
\Lambda_{ {\bf{k}} }({\bf{q}}))
\end{equation}
\[
\omega_{i}({\bf{q}})[d_{i}({\bf{q}}),a^{\dagger}_{ {\bf{k}} }({\bf{q}})]
 = {\tilde{\omega}}_{ {\bf{k}} }({\bf{q}})
[d_{i}({\bf{q}}),a^{\dagger}_{ {\bf{k}} }({\bf{q}})]
\]
\begin{equation}
- \Lambda_{ {\bf{k}} }(-{\bf{q}})
\sum_{ {\bf{k}} \neq {\bf{k}}^{'} }
\frac{v_{ {\bf{k}}-{\bf{k}}^{'} } }{V}
(\Lambda_{ {\bf{k}}^{'} }({\bf{q}})
[d_{i}({\bf{q}}),a_{ {\bf{k}}^{'} }(-{\bf{q}})]
 + [d_{i}({\bf{q}}),a^{\dagger}_{ {\bf{k}}^{'} }({\bf{q}})]
\Lambda_{ {\bf{k}}^{'} }(-{\bf{q}}))
\end{equation}
\[
\omega_{i}({\bf{q}})[d_{i}({\bf{q}}),a_{ {\bf{k}} }(-{\bf{q}})]
 = -{\tilde{\omega}}_{ {\bf{k}} }(-{\bf{q}})
[d_{i}({\bf{q}}),a_{ {\bf{k}} }(-{\bf{q}})]
\]
\begin{equation}
+ \Lambda_{ {\bf{k}} }({\bf{q}})
\sum_{ {\bf{k}} \neq {\bf{k}}^{'} }
\frac{v_{ {\bf{k}}-{\bf{k}}^{'} } }{V}
(\Lambda_{ {\bf{k}}^{'} }({\bf{q}})
[d_{i}({\bf{q}}),a_{ {\bf{k}}^{'} }(-{\bf{q}})]
 + [d_{i}({\bf{q}}),a^{\dagger}_{ {\bf{k}}^{'} }({\bf{q}})]
\Lambda_{ {\bf{k}}^{'} }(-{\bf{q}}))
\end{equation}
\begin{equation}
[d_{i}({\bf{q}}),a^{\dagger}_{ {\bf{k}} }({\bf{q}})]
 = -\frac{ \Lambda_{ {\bf{k}} }(-{\bf{q}}) }
{\omega_{i}({\bf{q}}) - {\tilde{\omega}}_{ {\bf{k}} }({\bf{q}}) }
\int \mbox{       }d{\vec{r}}\mbox{        }
e^{i{\bf{k}}.{\vec{r}}}v({\vec{r}})
\{R_{1}({\bf{q}},{\vec{r}}) + R_{2}({\bf{q}},{\vec{r}}) \}
\end{equation}
\begin{equation}
[d_{i}({\bf{q}}),a_{ {\bf{k}} }(-{\bf{q}})]
 = \frac{ \Lambda_{ {\bf{k}} }({\bf{q}}) }
{\omega_{i}({\bf{q}}) + {\tilde{\omega}}_{ {\bf{k}} }(-{\bf{q}}) }
\int 
\mbox{       }d{\vec{r}}\mbox{        }
e^{i{\bf{k}}.{\vec{r}}}v({\vec{r}})
\{R_{1}({\bf{q}},{\vec{r}}) + R_{2}({\bf{q}},{\vec{r}}) \}
\end{equation}
Here,
\begin{equation}
R_{1}({\bf{q}},{\vec{r}}) = \frac{1}{V}\sum_{ {\bf{k}}^{'} }
e^{-i{\bf{k}}^{'}.{\vec{r}} }
\Lambda_{ {\bf{k}}^{'} }({\bf{q}})
[d_{i}({\bf{q}}),a_{ {\bf{k}}^{'} }(-{\bf{q}})]
\end{equation}
\begin{equation}
R_{2}({\bf{q}},{\vec{r}}) = \frac{1}{V}\sum_{ {\bf{k}}^{'} }
e^{-i{\bf{k}}^{'}.{\vec{r}} }
\Lambda_{ {\bf{k}}^{'} }(-{\bf{q}})
[d_{i}({\bf{q}}),a^{\dagger}_{ {\bf{k}}^{'} }({\bf{q}})]
\end{equation}
\begin{equation}
(\omega_{i}({\bf{q}})
- {\tilde{\omega}}_{ i\nabla_{ {\vec{r}}^{'} } }({\bf{q}}) )
R_{2}({\bf{q}},{\vec{r}}^{'}) =
-\int \mbox{      }d{\vec{r}}\mbox{        }
F_{1}({\vec{r}}-{\vec{r}}^{'};{\bf{q}})
v({\vec{r}})\{R_{1}({\bf{q}},{\vec{r}}) + R_{2}({\bf{q}},{\vec{r}}) \}
\end{equation}
\begin{equation}
(\omega_{i}({\bf{q}})
- {\tilde{\omega}}_{ i\nabla_{ {\vec{r}}^{'} } }({\bf{q}}) )
R_{1}({\bf{q}},{\vec{r}}^{'}) =
\int \mbox{      }d{\vec{r}}\mbox{        }
F_{2}({\vec{r}}-{\vec{r}}^{'};{\bf{q}})
v({\vec{r}})\{R_{1}({\bf{q}},{\vec{r}}) + R_{2}({\bf{q}},{\vec{r}}) \}
\end{equation}
\begin{equation}
F_{1}({\vec{r}}-{\vec{r}}^{'};{\bf{q}})
 = \frac{1}{V}\sum_{ {\bf{k}} }
\Lambda_{ {\bf{k}} }(-{\bf{q}})
e^{i{\bf{k}}.({\vec{r}}-{\vec{r}}^{'})}
\end{equation}
\begin{equation}
F_{2}({\vec{r}}-{\vec{r}}^{'};{\bf{q}})
 = \frac{1}{V}\sum_{ {\bf{k}} }
\Lambda_{ {\bf{k}} }({\bf{q}})
e^{i{\bf{k}}.({\vec{r}}-{\vec{r}}^{'})}
\end{equation}
Now define,
\begin{equation}
R_{1}({\bf{q}},{\vec{r}}) + R_{2}({\bf{q}},{\vec{r}})
 = R({\bf{q}},{\vec{r}})
\end{equation}
\begin{equation}
(\omega_{i}({\bf{q}})
- {\tilde{\omega}}_{ i\nabla_{ {\vec{r}}^{'} } }({\bf{q}}) )
R({\bf{q}},{\vec{r}}^{'}) =
\int\mbox{             }d^{3}r\mbox{            }
 f_{0}({\bf{q}};{\vec{r}}-{\vec{r}}^{'})
v({\vec{r}})R({\bf{q}},{\vec{r}})
\end{equation}
\begin{equation}
f_{0}({\bf{q}};{\vec{r}}-{\vec{r}}^{'})
 = \int \mbox{           }\frac{ d^{3}k }{(2\pi)^{3}}\mbox{            }
n_{F}({\bf{k}})
e^{i{\bf{k}}.({\vec{r}}-{\vec{r}}^{'})}
(-2i)sin(\frac{1}{2}{\bf{q}}.({\vec{r}}-{\vec{r}}^{'}))
\end{equation}
\[
f_{0}({\bf{q}};{\vec{r}}-{\vec{r}}^{'})
 = \int_{0}^{k_{F}} \mbox{           }\frac{ 4\pi k \mbox{       }
dk}{(2\pi)^{3}}\mbox{            }
\frac{ sin(k|{\vec{r}}-{\vec{r}}^{'}|) }
{ |{\vec{r}}-{\vec{r}}^{'}| }
(-2i)sin(\frac{1}{2}{\bf{q}}.({\vec{r}}-{\vec{r}}^{'}))
\]
\begin{equation}
 = \frac{4\pi}{(2\pi)^{3}}
\frac{(-2i)sin(\frac{1}{2}{\bf{q}}.({\vec{r}}-{\vec{r}}^{'}))}
{|{\vec{r}}-{\vec{r}}^{'}|}
\{
-k_{F}\mbox{        }\frac{ cos(k_{F}|{\vec{r}}-{\vec{r}}^{'}|) }
{ |{\vec{r}}-{\vec{r}}^{'}| }
 +
\frac{ sin(k_{F}|{\vec{r}}-{\vec{r}}^{'}|) }
{ |{\vec{r}}-{\vec{r}}^{'}|^{2} } \}
 = i\mbox{       }C_{0}({\bf{q}}) {\bf{q}}.\nabla_{ {\vec{r}} }
\delta^{3}({\vec{r}} - {\vec{r}}^{'})
\end{equation}
\begin{equation}
(\omega_{i}({\bf{q}}) - i\frac{ {\bf{q}}.\nabla_{ {\vec{r}}^{'} } }{m})
R({\bf{q}},{\vec{r}}^{'})
 = -i\mbox{          }
C_{0}({\bf{q}})({\bf{q}}.\nabla_{ {\vec{r}}^{'} })v({\vec{r}}^{'})
R({\bf{q}},{\vec{r}}^{'})
\end{equation}
\begin{equation}
[ \omega_{i}({\bf{q}}) - i\frac{ {\bf{q}}.\nabla_{ {\vec{r}}^{'} } }{m}
+ i\mbox{          }
C_{0}({\bf{q}})({\bf{q}}.\nabla_{ {\vec{r}}^{'} }v({\vec{r}}^{'})) ]
R({\bf{q}},{\vec{r}}^{'})
 = -i\mbox{          }
C_{0}({\bf{q}})v({\vec{r}}^{'})
({\bf{q}}.\nabla_{ {\vec{r}}^{'} })R({\bf{q}},{\vec{r}}^{'})
\end{equation}
 This eigenvalue problem admits many eigenvalues. Actually, if not
 for the constraint of the square integrability of $ R({\bf{q}},{\vec{r}}) $,
 for each $ {\bf{q}} $ any positive real number would be an eigenvalue.
\begin{equation}
[ \omega_{i}({\bf{q}}) - i\frac{ {\bf{q}}.\nabla_{ {\vec{r}}^{'} } }{m}
+ i\mbox{          }
C_{0}({\bf{q}})({\bf{q}}.\nabla_{ {\vec{r}}^{'} }v({\vec{r}}^{'})) ]
R({\bf{q}},{\vec{r}}^{'})
 = -i\mbox{          }
C_{0}({\bf{q}})v({\vec{r}}^{'})
({\bf{q}}.\nabla_{ {\vec{r}}^{'} })R({\bf{q}},{\vec{r}}^{'})
\end{equation}
Let us now solve this equation,
\begin{equation}
R({\bf{q}},{\vec{r}}) = R({\bf{q}},{\vec{r}}_{0})
exp(\frac{1}{i\mbox{      }q\mbox{      }cos \theta}
\int^{r}_{ r_{0} }dr^{'}\frac{ \omega_{i}({\bf{q}})
+ i \mbox{          }C_{0}({\bf{q}})\mbox{      }q\mbox{     }
cos \theta \mbox{       }\frac{ \partial v(r^{'}) }{\partial r^{'}} }
{ \frac{1}{m} - C_{0}({\bf{q}})v(r^{'}) })
\end{equation}
For some suitable $ {\vec{r}}_{0} $.          
        
 This work was supported by the Dept. of Physics at
 University of Illinois at Urbana-Champaign. The authors may be contacted at
 the e-mail address setlur@mrlxpa.mrl.uiuc.edu.

\section{Appendix}

 In this appendix, we show how to derive the full dielectric function
 using the generalised RPA. Along the way we point out some pitfalls
 and possible generalisations.
 Let us write the generalised RPA hamiltonian and try and compute the
 dielectric function.
\begin{equation}
H_{0} = \sum_{ {\bf{k}} } \epsilon_{ {\bf{k}} }n_{0}( {\bf{k}} )
 -
\sum_{ {\bf{q}} \neq 0 }
\frac{ v_{ {\bf{q}} } }{2V}
 \sum_{ {\bf{k}} }n_{0}( {\bf{k}}+{\bf{q}}/2 )n_{0}( {\bf{k}}-{\bf{q}}/2 )
\end{equation}
\begin{equation}
H_{ext}(t) = \sum_{ {\bf{q}} }(U_{ext}({\bf{q}}t) + U^{*}_{ext}(-{\bf{q}}t))
\sum_{ {\bf{k}} }n_{ {\bf{q}} }({\bf{k}})
\end{equation}
here $ n_{ {\bf{q}} }({\bf{k}}) = c^{\dagger}_{ {\bf{k}} + {\bf{q}}/2 } c_{  {\\bf{k}}-{\bf{q}}/2  } $
and $ n_{ 0 }({\bf{k}}) =  c^{\dagger}_{ {\bf{k}} }c_{ {\bf{k}} } $
Let us now write down the quation of motion for $ n_{ {\bf{q}} }({\bf{k}}) $.
\[
i\frac{ \partial }{ \partial t}n_{ {\bf{q}} }({\bf{k}})
 = \sum_{ {\bf{k}}^{'} }\epsilon_{ {\bf{k}}^{'} }
[n_{ {\bf{q}} }({\bf{k}}), n_{0}( {\bf{k}}^{'} )]
\]
\[
 - \sum_{ {\bf{q}}^{'} \neq 0 }
\frac{ v_{ {\bf{q}}^{'} } }{V}
 \sum_{ {\bf{k}}^{'} }
[n_{ {\bf{q}} }({\bf{k}}),n_{0}( {\bf{k}}^{'})]
n_{0}( {\bf{k}}^{'}-{\bf{q}}^{'} )
\]
\begin{equation}
+ \sum_{ {\bf{k}}^{'},{\bf{q}}^{'} }
(U_{ext}({\bf{q}}^{'}t) + U^{*}_{ext}(-{\bf{q}}^{'}t))
[n_{ {\bf{q}} }({\bf{k}}),n_{ {\bf{q}}^{'} }( {\bf{k}}^{'})]
\end{equation}
Now,
\[
[n_{ {\bf{q}} }({\bf{k}}),n_{ {\bf{q}}^{'} }( {\bf{k}}^{'})]
 = [c^{\dagger}_{ {\bf{k}} + {\bf{q}}/2 }c_{ {\bf{k}} - {\bf{q}}/2 },
 c^{\dagger}_{ {\bf{k}}^{'} + {\bf{q}}^{'}/2 }
c_{ {\bf{k}}^{'} - {\bf{q}}^{'}/2 }]
\]
\begin{equation}
 = c^{\dagger}_{ {\bf{k}} + {\bf{q}}/2 }c_{ {\bf{k}}^{'} - {\bf{q}}^{'}/2 }
\delta_{ {\bf{k}} - {\bf{q}}/2, {\bf{k}}^{'} + {\bf{q}}^{'}/2 }
 - c^{\dagger}_{ {\bf{k}}^{'} + {\bf{q}}^{'}/2 }c_{ {\bf{k}} - {\bf{q}}/2 }
\delta_{ {\bf{k}} + {\bf{q}}/2, {\bf{k}}^{'} - {\bf{q}}^{'}/2 }
\end{equation}
One may approximate this as,
\[
[n_{ {\bf{q}} }({\bf{k}}),n_{ {\bf{q}}^{'} }( {\bf{k}}^{'})]
 = [c^{\dagger}_{ {\bf{k}} + {\bf{q}}/2 }c_{ {\bf{k}} - {\bf{q}}/2 },
 c^{\dagger}_{ {\bf{k}}^{'} + {\bf{q}}^{'}/2 }
c_{ {\bf{k}}^{'} - {\bf{q}}^{'}/2 }]
\]
\begin{equation}
 = [ n_{0}({\bf{k}}+{\bf{q}}/2) - n_{0}({\bf{k}}-{\bf{q}}/2) ]
\delta_{ {\bf{k}}, {\bf{k}}^{'} }\delta_{ {\bf{q}}, -{\bf{q}}^{'} }
\end{equation}
 The next approximation would be to replace the number operator by its c-number
 expectation value. But we shall desist from that for the moment.
\begin{equation}
[n_{ {\bf{q}} }({\bf{k}}), n_{0}( {\bf{k}}^{'} )]
 = n_{ {\bf{q}} }({\bf{k}})(\delta_{ {\bf{k}}^{'} , {\bf{k}}-{\bf{q}}/2 }
 - \delta_{ {\bf{k}}^{'} , {\bf{k}}+{\bf{q}}/2 })
\end{equation}
\[
i\frac{ \partial }{ \partial t}n_{ {\bf{q}} }({\bf{k}})
 = (\epsilon_{ {\bf{k}}-{\bf{q}}/2 }-\epsilon_{ {\bf{k}}+{\bf{q}}/2 })
 n_{ {\bf{q}} }({\bf{k}})
 - \sum_{ {\bf{q}}^{'} \neq 0 }
\frac{ v_{ {\bf{q}}^{'} } }{V}
(n_{0}( {\bf{k}}-{\bf{q}}/2-{\bf{q}}^{'} )
- n_{0}({\bf{k}}+{\bf{q}}/2-{\bf{q}}^{'} ))
 n_{ {\bf{q}} }({\bf{k}})
\]
\begin{equation}
+
(U_{ext}(-{\bf{q}}t) + U^{*}_{ext}({\bf{q}}t))
(n_{0}({\bf{k}}+{\bf{q}}/2) - n_{0}({\bf{k}}-{\bf{q}}/2))
\end{equation}
\begin{equation}
\langle n_{ {\bf{q}} }({\bf{k}}) \rangle
 = U_{ext}(-{\bf{q}}t)C_{ {\bf{q}} }({\bf{k}})
+  U^{*}_{ext}({\bf{q}}t)D_{ {\bf{q}} }({\bf{k}})
\end{equation}
Let us now ignore fluctuations in the momentum distribution. That is,
 we are allowed to replace
\begin{equation}
\langle n_{ 0 }({\bf{k}}^{'}) n_{ {\bf{q}} }({\bf{k}}) \rangle
 = \langle n_{ 0 }({\bf{k}}^{'}) \rangle\langle n_{ {\bf{q}} }({\bf{k}}) \rangle
\end{equation}
\[
\omega\mbox{      } U_{ext}(-{\bf{q}}t)C_{ {\bf{q}} }({\bf{k}})
- \omega\mbox{      }U^{*}_{ext}({\bf{q}}t)D_{ {\bf{q}} }({\bf{k}})
 =  (\epsilon_{ {\bf{k}}-{\bf{q}}/2 }-\epsilon_{ {\bf{k}}+{\bf{q}}/2 })
 U_{ext}(-{\bf{q}}t)C_{ {\bf{q}} }({\bf{k}})
+  (\epsilon_{ {\bf{k}}-{\bf{q}}/2 }-\epsilon_{ {\bf{k}}+{\bf{q}}/2 })
 U^{*}_{ext}({\bf{q}}t)D_{ {\bf{q}} }({\bf{k}})
\]
\[
 - \sum_{ {\bf{q}}^{'} \neq 0 }
\frac{ v_{ {\bf{q}}^{'} } }{V}
(\langle n_{0}( {\bf{k}}-{\bf{q}}/2-{\bf{q}}^{'} ) \rangle
- \langle n_{0}({\bf{k}}+{\bf{q}}/2-{\bf{q}}^{'} ) \rangle )
  U_{ext}(-{\bf{q}}t)C_{ {\bf{q}} }({\bf{k}})
\]
\[
- \sum_{ {\bf{q}}^{'} \neq 0 }
\frac{ v_{ {\bf{q}}^{'} } }{V}
(\langle n_{0}( {\bf{k}}-{\bf{q}}/2-{\bf{q}}^{'} ) \rangle
- \langle n_{0}({\bf{k}}+{\bf{q}}/2-{\bf{q}}^{'} ) \rangle )
  U^{*}_{ext}({\bf{q}}t)D_{ {\bf{q}} }({\bf{k}})
\]
\begin{equation}
+
(U_{ext}(-{\bf{q}}t) + U^{*}_{ext}({\bf{q}}t))
(\langle n_{0}({\bf{k}}+{\bf{q}}/2) \rangle
 - \langle n_{0}({\bf{k}}-{\bf{q}}/2) \rangle)
\end{equation}
\[
\omega\mbox{      }C_{ {\bf{q}} }({\bf{k}})
 = (\epsilon_{ {\bf{k}}-{\bf{q}}/2 }-\epsilon_{ {\bf{k}}+{\bf{q}}/2 })
C_{ {\bf{q}} }({\bf{k}})- \sum_{ {\bf{q}}^{'} \neq 0 }
\frac{ v_{ {\bf{q}}^{'} } }{V}
(\langle n_{0}( {\bf{k}}-{\bf{q}}/2-{\bf{q}}^{'} ) \rangle
- \langle n_{0}({\bf{k}}+{\bf{q}}/2-{\bf{q}}^{'} ) \rangle )
 C_{ {\bf{q}} }({\bf{k}})
\]
\begin{equation}
+
(\langle n_{0}({\bf{k}}+{\bf{q}}/2) \rangle
 - \langle n_{0}({\bf{k}}-{\bf{q}}/2) \rangle)
\end{equation}
Since,
\begin{equation}
U_{eff}({\bf{q}}t) = U_{ext}({\bf{q}}t) + \frac{ v_{ {\bf{q}} } }{V}
\langle \rho^{'}_{ -{\bf{q}} } \rangle U_{ext}({\bf{q}}t)
\end{equation}
\begin{equation}
\langle \rho^{'}_{ -{\bf{q}} } \rangle
 = \sum_{ {\bf{k}} }C_{ -{\bf{q}} }({\bf{k}})
\end{equation}
\begin{equation}
 \sum_{ {\bf{k}} }C_{ -{\bf{q}} }({\bf{k}})
 = \sum_{ {\bf{k}} }
\frac{ \langle n_{0}({\bf{k}}-{\bf{q}}/2) \rangle
 - \langle n_{0}({\bf{k}}+{\bf{q}}/2) \rangle }
{\omega - {\tilde{\epsilon}}_{ {\bf{k}} + {\bf{q}}/2 }
 + {\tilde{\epsilon}}_{ {\bf{k}} - {\bf{q}}/2 } }
\end{equation}
If we use the definition of the dielectric function we get a wrong answer.
\begin{equation}
\epsilon_{WRONG}({\bf{q}}, \omega) =
\frac{ U_{ext}({\bf{q}}t) }{  U_{eff}({\bf{q}}t) }
 = \frac{1}{ 1 + \frac{ v_{ {\bf{q}} } }{V}
\sum_{ {\bf{k}} }
\frac{ \langle n_{0}({\bf{k}}-{\bf{q}}/2) \rangle
 - \langle n_{0}({\bf{k}}+{\bf{q}}/2) \rangle }
{\omega - {\tilde{\epsilon}}_{ {\bf{k}} + {\bf{q}}/2 }
 + {\tilde{\epsilon}}_{ {\bf{k}} - {\bf{q}}/2 } } }
\end{equation}
 The reason is because we have been very cavalier in our treatment of
 correlations. 
 One has to consider the full Hamiltonian when dealing
 with the dielectric function rather than just $ H_{0} $.
 Let us now try and do this.
 \[
i\frac{ \partial }{ \partial t} n^{t}_{ {\bf{q}} }({\bf{k}})
 = (\epsilon_{ {\bf{k}}-{\bf{q}}/2 } - \epsilon_{ {\bf{k}}+{\bf{q}}/2 })
 n^{t}_{ {\bf{q}} }({\bf{k}})
+ \sum_{ {\bf{q}}^{'} \neq 0 }\frac{ v_{ {\bf{q}}^{'} } }{2V}
[n_{ {\bf{q}} }({\bf{k}}), \rho_{ {\bf{q}}^{'} }] \rho^{t}_{ -{\bf{q}}^{'} }
+ \sum_{ {\bf{q}}^{'} \neq 0 }\frac{ v_{ {\bf{q}}^{'} } }{2V}
\rho^{t}_{ {\bf{q}}^{'} }[n_{ {\bf{q}} }({\bf{k}}), \rho_{ -{\bf{q}}^{'} }]
\]
\begin{equation}
 + \sum_{ {\bf{q}}^{'} \neq 0 }(U_{ext}({\bf{q}}^{'}t)
+ U^{*}_{ext}(-{\bf{q}}^{'}t))
[n_{ {\bf{q}} }({\bf{k}}),\rho_{ {\bf{q}}^{'} }]
\end{equation}
\[
[n_{ {\bf{q}} }({\bf{k}}),\rho_{ {\bf{q}}^{'} }]
 = \sum_{ {\bf{k}}^{'} }
[c^{\dagger}_{ {\bf{k}} + {\bf{q}}/2 } c_{  {\bf{k}} - {\bf{q}}/2 },
c^{\dagger}_{ {\bf{k}}^{'} + {\bf{q}}^{'}/2 }
 c_{  {\bf{k}}^{'} - {\bf{q}}^{'}/2 }]
\]
\[
 = \sum_{ {\bf{k}}^{'} }
c^{\dagger}_{ {\bf{k}} + {\bf{q}}/2 }c_{  {\bf{k}}^{'} - {\bf{q}}^{'}/2 }
\delta_{ {\bf{k}} - {\bf{q}}/2, {\bf{k}}^{'} + {\bf{q}}^{'}/2 }
 - \sum_{ {\bf{k}}^{'} }
c^{\dagger}_{ {\bf{k}}^{'} + {\bf{q}}^{'}/2 }c_{  {\bf{k}} - {\bf{q}}/2 }
\delta_{ {\bf{k}} + {\bf{q}}/2, {\bf{k}}^{'} - {\bf{q}}^{'}/2 }
\]
\begin{equation}
\approx \delta_{ {\bf{q}}, -{\bf{q}}^{'} }
[n_{0}({\bf{k}} + {\bf{q}}/2) - n_{0}({\bf{k}} - {\bf{q}}/2)]
\end{equation}
\[
i\frac{ \partial }{ \partial t} n^{t}_{ {\bf{q}} }({\bf{k}})
 = (\epsilon_{ {\bf{k}}-{\bf{q}}/2 } - \epsilon_{ {\bf{k}}+{\bf{q}}/2 })
 n^{t}_{ {\bf{q}} }({\bf{k}})
+ \frac{ v_{ {\bf{q}} } }{V}
(n_{0}({\bf{k}} + {\bf{q}}/2) - n_{0}({\bf{k}} - {\bf{q}}/2))
 \rho^{t}_{ {\bf{q}} }
\]
\begin{equation}
 + (U_{ext}(-{\bf{q}}t)
+ U^{*}_{ext}({\bf{q}}t))
(n_{0}({\bf{k}} + {\bf{q}}/2) - n_{0}({\bf{k}} - {\bf{q}}/2))
\end{equation}
 Let us make a first pass at the computation of the dielectric function.
 Here, we make use of mean-field theory, that is, replace
\begin{equation}
 \langle n_{0}( {\bf{k}}^{'} )\rho_{ {\bf{q}} } \rangle =
 \langle n_{0}( {\bf{k}}^{'} )\rangle \langle \rho_{ {\bf{q}} } \rangle 
\end{equation}
\[
\omega \mbox{      }\langle  n^{t}_{ {\bf{q}} }({\bf{k}}) \rangle
 = (\epsilon_{ {\bf{k}}-{\bf{q}}/2 } - \epsilon_{ {\bf{k}}+{\bf{q}}/2 })
\langle  n^{t}_{ {\bf{q}} }({\bf{k}}) \rangle
+ \frac{ v_{ {\bf{q}} } }{V}
(\langle  n_{0}({\bf{k}} + {\bf{q}}/2)  \rangle
 -  \langle  n_{0}({\bf{k}} - {\bf{q}}/2)  \rangle)
\langle  \rho^{t}_{ {\bf{q}} } \rangle
\]
\begin{equation}
 + (U_{ext}(-{\bf{q}}t)
+ U^{*}_{ext}({\bf{q}}t))
(\langle  n_{0}({\bf{k}} + {\bf{q}}/2) \rangle -
\langle  n_{0}({\bf{k}} - {\bf{q}}/2) \rangle)
\end{equation}
\[
\langle  n^{t}_{ {\bf{q}} }({\bf{k}}) \rangle
 = \frac{ v_{ {\bf{q}} } }{V}
\frac{\langle  n_{0}({\bf{k}} + {\bf{q}}/2)  \rangle
 -  \langle  n_{0}({\bf{k}} - {\bf{q}}/2)  \rangle }
{ \omega  - \epsilon_{ {\bf{k}}-{\bf{q}}/2 } +
 \epsilon_{ {\bf{k}}+{\bf{q}}/2 }  }\langle  \rho^{t}_{ {\bf{q}} } \rangle
\]
\begin{equation}
 + (U_{ext}(-{\bf{q}}t)
+ U^{*}_{ext}({\bf{q}}t))
\frac{\langle  n_{0}({\bf{k}} + {\bf{q}}/2)  \rangle
 -  \langle  n_{0}({\bf{k}} - {\bf{q}}/2)  \rangle }
{ \omega  - \epsilon_{ {\bf{k}}-{\bf{q}}/2 }  +\epsilon_{ {\bf{k}}+{\bf{q}}/2 }}
\end{equation}
This means,
\begin{equation}
\langle \rho_{ -{\bf{q}} } \rangle = U_{ext}({\bf{q}}t)
\frac{ P_{0}({\bf{q}},\omega) }{ \epsilon({\bf{q}}, \omega) }
\end{equation}
\begin{equation}
 P_{0}({\bf{q}},\omega) = \sum_{ {\bf{k}} }
\frac{ \langle  n_{0}({\bf{k}} - {\bf{q}}/2)  \rangle -  \langle  n_{0}({\bf{k}} + {\bf{q}}/2)  \rangle }
{\omega - \epsilon_{ {\bf{k}} + {\bf{q}}/2 }+ \epsilon_{ {\bf{k}} - {\bf{q}}/2 } }
\end{equation}
\begin{equation}
\epsilon({\bf{q}}, \omega) = 1 - \frac{v_{ {\bf{q}} }}{V}
P_{0}({\bf{q}},\omega)
\end{equation}
From this and the fact that
\begin{equation}
\epsilon_{g-RPA}({\bf{q}},\omega) = \frac{ U_{ext}({\bf{q}}t) }
{  U_{eff}({\bf{q}}t) } = \epsilon({\bf{q}},\omega)
\end{equation}
Next we would like to include fluctuations. Let us do this differently
 this time via the use of the BBGKY heirarchy.
\[
i\frac{ \partial }{ \partial t} \langle n^{t}_{ {\bf{q}} }({\bf{k}}) \rangle
 = (\epsilon_{ {\bf{k}}-{\bf{q}}/2 } - \epsilon_{ {\bf{k}}+{\bf{q}}/2 })
 \langle n^{t}_{ {\bf{q}} }({\bf{k}})  \rangle
+ \frac{ v_{ {\bf{q}} } }{V}
(\langle n_{0}({\bf{k}} + {\bf{q}}/2)  \rangle -
 \langle n_{0}({\bf{k}} - {\bf{q}}/2)  \rangle)
\langle  \rho^{t}_{ {\bf{q}} } \rangle
\]
\[
+  \frac{ v_{ {\bf{q}} } }{V}
(F_{2A}({\bf{k}} + {\bf{q}}/2, {\bf{q}})
- F_{2A}({\bf{k}} - {\bf{q}}/2, {\bf{q}}))
\]
\begin{equation}
 + (U_{ext}(-{\bf{q}}t)
+ U^{*}_{ext}({\bf{q}}t))
(n_{0}({\bf{k}} + {\bf{q}}/2) - n_{0}({\bf{k}} - {\bf{q}}/2))
\end{equation}
Here,
\begin{equation}
F_{2A}({\bf{k}}^{'},{\bf{q}};t) = \langle n_{0}({\bf{k}}^{'})
\rho^{t}_{ {\bf{q}} } \rangle - \langle n_{0}({\bf{k}}^{'})\rangle
 \langle \rho^{t}_{ {\bf{q}} } \rangle
\end{equation}
\begin{equation}
F_{2}({\bf{k}}^{'};{\bf{k}},{\bf{q}};t)
 = \langle n_{0}({\bf{k}}^{'})n^{t}_{ {\bf{q}} }({\bf{k}}) \rangle
 - \langle n_{0}({\bf{k}}^{'}) \rangle
 \langle n^{t}_{ {\bf{q}} }({\bf{k}}) \rangle
\end{equation}
\begin{equation}
F_{2A}({\bf{k}}^{'},{\bf{q}};t) = \sum_{ {\bf{k}} }
F_{2}({\bf{k}}^{'};{\bf{k}},{\bf{q}};t)
\end{equation}
\[
i\frac{ \partial }{ \partial t }F_{2}({\bf{k}}^{'}; {\bf{k}},{\bf{q}};t)
 = (\epsilon_{ {\bf{k}}-{\bf{q}}/2 } - \epsilon_{ {\bf{k}}+{\bf{q}}/2 })
F_{2}({\bf{k}}^{'}; {\bf{k}},{\bf{q}};t)
 + \frac{ v_{ {\bf{q}} } }{V}(N({\bf{k}}^{'},{\bf{k}}+{\bf{q}}/2)
 - N({\bf{k}}^{'},{\bf{k}}-{\bf{q}}/2))
\langle \rho^{t}_{ {\bf{q}} } \rangle
\]
\begin{equation}
+  \frac{ v_{ {\bf{q}} } }{V}(\langle n_{0}({\bf{k}}+{\bf{q}}/2) \rangle
 - \langle n_{0}({\bf{k}}-{\bf{q}}/2) \rangle)
F_{2A}({\bf{k}}^{'},{\bf{q}};t)
 + (U_{ext}(-{\bf{q}}t) + U^{*}_{ext}({\bf{q}}t))
(N({\bf{k}}^{'}, {\bf{k}}+{\bf{q}}/2) - N({\bf{k}}^{'}, {\bf{k}}-{\bf{q}}/2))
\end{equation}
Let us write,
\begin{equation}
F_{2}({\bf{k}}^{'}; {\bf{k}},{\bf{q}};t) = U_{ext}(-{\bf{q}}t)
F_{2,a}({\bf{k}}^{'}; {\bf{k}},{\bf{q}})
 + U^{*}_{ext}({\bf{q}}t) F_{2,b}({\bf{k}}^{'}; {\bf{k}},{\bf{q}})
\end{equation}
\begin{equation}
\langle \rho^{t}_{ {\bf{q}} } \rangle = U_{ext}(-{\bf{q}}t)
\langle \rho^{'}_{ {\bf{q}} } \rangle
+  U^{*}_{ext}({\bf{q}}t) \langle \rho^{''}_{ {\bf{q}} } \rangle
\end{equation}
Also define,
\begin{equation}
N({\bf{k}},{\bf{k}}^{'}) = \langle n_{0}({\bf{k}})n_{0}({\bf{k}}^{'}) \rangle
 - \langle n_{0}({\bf{k}}) \rangle \langle  n_{0}({\bf{k}}^{'}) \rangle
\end{equation}
\[
\omega \mbox{      }F_{2,a}({\bf{k}}^{'}; {\bf{k}},{\bf{q}})
 = (\epsilon_{ {\bf{k}}-{\bf{q}}/2 } - \epsilon_{ {\bf{k}}+{\bf{q}}/2 })
F_{2,a}({\bf{k}}^{'}; {\bf{k}},{\bf{q}})
 + \frac{ v_{ {\bf{q}} } }{V}(N({\bf{k}}^{'},{\bf{k}}+{\bf{q}}/2)
 - N({\bf{k}}^{'},{\bf{k}}-{\bf{q}}/2))
\langle \rho^{'}_{ {\bf{q}} } \rangle
\]
\begin{equation}
+  \frac{ v_{ {\bf{q}} } }{V}(\langle n_{0}({\bf{k}}+{\bf{q}}/2) \rangle
 - \langle n_{0}({\bf{k}}-{\bf{q}}/2) \rangle)
F^{a}_{2A}({\bf{k}}^{'},{\bf{q}})
 +
(N({\bf{k}}^{'}, {\bf{k}}+{\bf{q}}/2) - N({\bf{k}}^{'}, {\bf{k}}-{\bf{q}}/2))
\end{equation}
\[
-\omega \mbox{      }F_{2,b}({\bf{k}}^{'}; {\bf{k}},{\bf{q}})
 = (\epsilon_{ {\bf{k}}-{\bf{q}}/2 } - \epsilon_{ {\bf{k}}+{\bf{q}}/2 })
F_{2,b}({\bf{k}}^{'}; {\bf{k}},{\bf{q}})
 + \frac{ v_{ {\bf{q}} } }{V}(N({\bf{k}}^{'},{\bf{k}}+{\bf{q}}/2)
 - N({\bf{k}}^{'},{\bf{k}}-{\bf{q}}/2))
\langle \rho^{''}_{ {\bf{q}} } \rangle
\]
\begin{equation}
+  \frac{ v_{ {\bf{q}} } }{V}(\langle n_{0}({\bf{k}}+{\bf{q}}/2) \rangle
 - \langle n_{0}({\bf{k}}-{\bf{q}}/2) \rangle)
F^{b}_{2A}({\bf{k}}^{'},{\bf{q}})
 +
(N({\bf{k}}^{'}, {\bf{k}}+{\bf{q}}/2) - N({\bf{k}}^{'}, {\bf{k}}-{\bf{q}}/2))
\end{equation}
\[
\epsilon({\bf{q}},\omega) \mbox{         }F_{2A}^{a}({\bf{k}}^{'},{\bf{q}})
 = \frac{ v_{ {\bf{q}} } }{V}\sum_{ {\bf{k}} }
\frac{ N({\bf{k}}^{'}, {\bf{k}}+{\bf{q}}/2)
 -  N({\bf{k}}^{'}, {\bf{k}}-{\bf{q}}/2) }
{\omega - \epsilon_{ {\bf{k}}-{\bf{q}}/2 } +  \epsilon_{ {\bf{k}}+{\bf{q}}/2 }}
\langle \rho^{'}_{ {\bf{q}} } \rangle
\]
\begin{equation}
+ \sum_{ {\bf{k}} }\frac{ N({\bf{k}}^{'}, {\bf{k}}+{\bf{q}}/2)
 -  N({\bf{k}}^{'}, {\bf{k}}-{\bf{q}}/2) }
{\omega - \epsilon_{ {\bf{k}}-{\bf{q}}/2 } +  \epsilon_{ {\bf{k}}+{\bf{q}}/2 }}
\end{equation}
\[
\omega\mbox{   }C_{ {\bf{q}} }({\bf{k}})
 = (\epsilon_{ {\bf{k}}-{\bf{q}}/2 } - \epsilon_{ {\bf{k}}+{\bf{q}}/2 })
C_{ {\bf{q}} }({\bf{k}})
+ \frac{ v_{ {\bf{q}} } }{V}(\langle n_{0}({\bf{k}}+{\bf{q}}/2) \rangle
 - \langle n_{0}({\bf{k}}-{\bf{q}}/2) \rangle)
\langle \rho^{'}_{ {\bf{q}} } \rangle
+  \frac{ v_{ {\bf{q}} } }{V}(F^{a}_{2A}({\bf{k}}+{\bf{q}}/2)
 - F^{a}_{2A}({\bf{k}}-{\bf{q}}/2))
\]
\begin{equation}
+ (\langle n_{0}({\bf{k}}+{\bf{q}}/2) \rangle
- \langle n_{0}({\bf{k}}-{\bf{q}}/2) \rangle)
\end{equation}
\[
C_{ {\bf{q}} }({\bf{k}}) = \frac{ v_{ {\bf{q}} } }{V}
\frac{ \langle n_{0}({\bf{k}}+{\bf{q}}/2) \rangle
 - \langle n_{0}({\bf{k}}-{\bf{q}}/2) \rangle }
{ \omega - \epsilon_{ {\bf{k}}-{\bf{q}}/2 } + \epsilon_{ {\bf{k}}+{\bf{q}}/2 } }
\langle \rho^{'}_{ {\bf{q}} } \rangle
\]
\[
+ \frac{ \langle n_{0}({\bf{k}}+{\bf{q}}/2) \rangle
 - \langle n_{0}({\bf{k}}-{\bf{q}}/2) \rangle }
{ \omega - \epsilon_{ {\bf{k}}-{\bf{q}}/2 } + \epsilon_{ {\bf{k}}+{\bf{q}}/2 } }
\]
\begin{equation}
+  \frac{ v_{ {\bf{q}} } }{V}
\frac{ F^{a}_{2A}({\bf{k}}+{\bf{q}}/2,{\bf{q}})
 - F^{a}_{2A}({\bf{k}}-{\bf{q}}/2,{\bf{q}}) }
{ \omega - \epsilon_{ {\bf{k}}-{\bf{q}}/2 } + \epsilon_{ {\bf{k}}+{\bf{q}}/2 } }
\end{equation}
 After all this, it may be shown that the overall dielectric function including
 possible fluctutations in the momentum distribution is given by,
\begin{equation}
\epsilon_{eff}({\bf{q}},\omega) = \epsilon_{g-RPA}({\bf{q}},\omega)
 - (\frac{v_{ {\bf{q}} } }{V})^{2}\frac{ P_{2}({\bf{q}},\omega) }
{ \epsilon_{g-RPA}({\bf{q}},\omega) }
\end{equation}
Here,
\begin{equation}
P_{2}({\bf{q}},\omega) = \sum_{ {\bf{k}},{\bf{k}}^{'} }
\frac{ N({\bf{k}}+{\bf{q}}/2, {\bf{k}}^{'}+{\bf{q}}/2)
- N({\bf{k}}-{\bf{q}}/2, {\bf{k}}^{'}+{\bf{q}}/2)
- N({\bf{k}}+{\bf{q}}/2, {\bf{k}}^{'}-{\bf{q}}/2)
+ N({\bf{k}}-{\bf{q}}/2, {\bf{k}}^{'}-{\bf{q}}/2) }
{ (\omega - \epsilon_{ {\bf{k}}-{\bf{q}}/2 } +\epsilon_{ {\bf{k}}+{\bf{q}}/2 })
(\omega
 - \epsilon_{ {\bf{k}}^{'}-{\bf{q}}/2 } +\epsilon_{ {\bf{k}}^{'}+{\bf{q}}/2 }) }
\end{equation}
\begin{equation}
\epsilon_{g-RPA}({\bf{q}},\omega) = 1 + \frac{ v_{ {\bf{q}} } }{V}
\sum_{ {\bf{k}} } \frac{ \langle n_{0}({\bf{k}}+{\bf{q}}/2) \rangle
 - \langle n_{0}({\bf{k}}-{\bf{q}}/2) \rangle }
{ \omega
 - \epsilon_{ {\bf{k}}+{\bf{q}}/2 } +\epsilon_{ {\bf{k}}-{\bf{q}}/2  } }
\end{equation}


\begin{thebibliography}{30}
\bibitem[12]{footnote} The pathological exception is for nonideal 
 distributions such as,
 $ \langle n_{ {\bf{k}} } \rangle = 0 $ for $ |{\bf{k}}| < 0.5\mbox{   }k_{f} $
 and $ \langle n_{ {\bf{k}} } \rangle = 1 $ for
 $ 0.5\mbox{   }k_{f} < |{\bf{k}}| < 1.5\mbox{   }k_{f} $
 and $ \langle n_{ {\bf{k}} } \rangle = 0 $ for
 $ |{\bf{k}}| > 1.5\mbox{   }k_{f} $ 
 this is very nonideal but has zero fluctuation, 
 but hopefully these are not physical.
\end{thebibliography}
\end{document}